\documentstyle[11pt,psfig]{article}
\def\baselinestretch {1.5}
\evensidemargin 10mm\oddsidemargin 10mm\def\cite#1{#1}
\newcommand{\ct}[1]{[\cite{#1}]}
\def\thebibliography#1{\section*{References}\list
 {[\arabic{enumi}]}{\settowidth\labelwidth{[#1]}\leftmargin\labelwidth
 \advance\leftmargin\labelsep
 \usecounter{enumi}}
 \def\newblock{\hskip .11em plus .33em minus -.07em}
 \sloppy
 \sfcode`\.=1000\relax}

\begin{document}
\begin{center}
{\Large\bf NON-STANDARD MODEL OF THE NUCLEON ELECTROMAGNETIC STRUCTURE
 AND ITS PREDICTABILITY}
\vspace{1mm}
\end{center}
\vspace{0.1cm}

\def\baselinestretch {1}
\begin{center}

{\bf S.Dubni\v cka}
\footnote{E-mail address: fyzidubn@nic.savba.sk}
\\[3mm]
{\em Int'l Centre for Theoretical Physics, Strada Costiera 11,
I-34014 Trieste, Italy\\
and \\
Institute of Physics, Slovak Academy of Sciences,
D\'ubravsk\'a cesta 9, 842 28 Bratislava, Slovak Republic.}\\

\vspace*{0.1cm}

{\bf A.Z.Dubni\v ckov\'a}
\footnote{E-mail address: dubnickova@fmph.uniba.sk}
and {\bf P. Weisenpacher}\\
{\em Department of Theoretical Physics, Comenius University, Mlynsk\'a dolina, 842 48 Bratislava, Slovak
Republic.}\\

\vspace{0.1cm}
\end{center}

\def\baselinestretch {1.5}
\def\sa{\omega}
\def\sbb{{\omega^,}}
\def\sc{{\omega^{,,}}}
\def\sd{{\phi}}
\def\se{{\phi^,}}
\def\va{{\varrho}}
\def\vb{{\varrho^,}}
\def\vc{{\varrho^{,,}}}
\def\vd{{\varrho^{,,,}}}
\def\ve{{\varrho^{,,,,}}}

\def\sam{\omega_0}
\def\sbbm{{\omega^,_0}}
\def\scm{{\omega^{,,}_0}}
\def\sdm{{\phi_0}}
\def\sem{{\phi^,_0}}
\def\vam{{\varrho_0}}
\def\vbm{{\varrho^,_0}}
\def\vcm{{\varrho^{,,}_0}}
\def\vdm{{\varrho^{,,,}_0}}
\def\vem{{\varrho^{,,,,}_0}}

\def\mmm#1{\frac{m_{#1}^2}{m_{#1}^2-t}}
\def\mmt#1#2{\frac{m_{#1}^2m_{#2}^2}{(m_{#1}^2-t)(m_{#2}^2-t)}}
\def\mtt#1#2#3{\frac{m_{#1}^2m_{#2}^2m_{#3}^2}{(m_{#1}^2-t)(m_{#2}^2-t)
     (m_{#3}^2-t)}}
\def\mc#1{{m^2_#1}}
\def\zll#1#2#3#4{{\frac{#1-#2}{#3-#4}}}
\def\zlll#1#2#3#4#5#6#7#8{{\frac{(#1-#2)(#3-#4)}{(#5-#6)(#7-#8)}}}
\def\zzl#1#2#3#4#5#6{{\frac{#1#2}{(#3-#4)(#5-#6)}}}
\def\xxx#1#2{\frac{(#1_N-#1_#2)(#1_N-#1_#2^*)(#1_N-1/#1_#2)(#1_N-1/#1_#2^*)}
               {(#1-#1_#2)(#1-#1_#2^*)(#1-1/#1_#2)(#1-1/#1_#2^*)}}
\def\eee#1#2{\frac{(#1_N-#1_#2)(#1_N-#1_#2^*)(#1_N+#1_#2)(#1_N+#1_#2^*)}
               {(#1-#1_#2)(#1-#1_#2^*)(#1+#1_#2)(#1+#1_#2^*)}}
\def\cc#1{{C_#1}}
\def\zl#1#2#3#4#5{{\frac{#2^#1-#3^#1}{#4^#1-#5^#1}}}
\def\uvv#1{{(\frac{1-#1^2}{1-#1_N^2})}}
\def\fff#1#2{(f^{(#1)}_{#2{NN}}/f_{#2})}
\def\ccc#1#2#3#4#5{\frac{C^{#1}_#2-C^{#1}_#3}{C^{#1}_#4-C^{#1}_#5}}

\def\mcsa{{\mc\sa}}
\def\mcsb{{\mc\sbb}}
\def\mcsd{{\mc\sd}}
\def\mcva{{\mc\va}}
\def\mcvb{{\mc\vb}}
\def\mcvc{{\mc\vc}}
\def\mcsc{{\mc\sc}}
\def\mcse{{\mc\se}}
\def\mcvd{{\mc\vd}}
\def\mcve{{\mc\ve}}

\def\ccsa{{\cc\sa}}
\def\ccsb{{\cc\sbb}}
\def\ccsd{{\cc\sd}}
\def\ccva{{\cc\va}}
\def\ccvb{{\cc\vb}}
\def\ccvc{{\cc\vc}}
\def\ccsc{{\cc\sc}}
\def\ccse{{\cc\se}}
\def\ccvd{{\cc\vd}}
\def\ccve{{\cc\ve}}

\begin{abstract}

Unitary and analytic ten-resonance model of the nucleon electromagnetic (e.m.)
structure with canonical normalizations and asymptotics is constructed on
a four-sheeted Riemann surface.
It describes well all existing experimental
space-like and time-like data on the nucleon e.m. form factors (ff's),
including also FENICE (Frascati) results on the neutron, for the first time.
This is achieved without any external constraints on the isovector spectral
functions following from the $\pi N$-scattering data and pion e.m. ff
behaviour through the unitarity
condition. Just opposite, the model itself predicts a pronounced effect of the two-pion
continuum on the isovector spectral functions revealing the strong enhancement
of the left wing of the $\rho(770)$-resonance close to two-pion threshold.

The existence of the fourth excited state
of the $\rho(770)$ meson with parameters
$m_{\rho^{,,,,}}=2506\pm 38$ MeV, $\Gamma_{\rho^{,,,,}}=700\pm 179$ MeV, the
large values of $f^{(1,2)}_{\phi NN}$ coupling constants,
indicating a violation of the OZI rule and also  the isoscalar
spectral function behaviours are predicted by the presented model.

\end{abstract}

\section{Introduction}

The electromagnetic  structure of (e.m.) the nucleons, as revealed in
elastic electron-nucleon scattering, is completely described by four
independent scalar functions, called form factors (ff's) and dependent on
the square momentum transfer $t=-Q^2$ of the virtual photon. They can be
chosen in a different way, e.g. as the Dirac and Pauli ff's, $F^p_1(t)$,
$F^n_1(t)$ and $F^p_2(t)$, $F^n_2(t)$, or the Sachs electric and magnetic
ff's, $G^p_E(t)$, $G^n_E(t)$ and $G_M^p(t)$, $G^n_M(t)$, or isoscalar and
isovector Dirac and Pauli ff's, $F_1^s(t)$, $F_1^v(t)$ and $F_2^s(t)$,
$F^v_2(t)$ and isoscalar and isovector electric and magnetic ff's,
$G_E^s(t)$, $G_E^v(t)$ and $G^s_M(t)$, $G^v_M(t)$, respectively.

The Dirac and Pauli ff's are naturally obtained in a decomposition of the
nucleon matrix element of the e.m. current into maximally linearly
independent covariants constructed from the four-momenta, $\gamma$-matrices
and Dirac bispinors of nucleons as follows
\begin{equation}
\langle N|J^{e.m.}_\mu|N\rangle=e\bar u(p')\{\gamma_\mu
F^N_1(t)+\frac{i}{2m_N}\sigma_{\mu\nu}(p'-p)_{\nu}F^N_2(t)\}u(p)\label{d1}
\end{equation}
with $m_N$ to be  nucleon mass.

On the other hand, the electric and magnetic ff's are very suitable in an
extraction of the experimental information on the nucleon e.m. structure
from the measured cross sections
\begin{equation}
\frac{d\sigma^{lab}(e^-N\to
e^-N)}{d\Omega}=\frac{\alpha^2}{4E^2}\frac{\cos^2(\theta/2)}{\sin^4(\theta/2)}
\frac{1}{1+(\frac{2E}{m_N})\sin^2(\theta/2)}[\frac{G^2_E-\frac{t}{4m_N^2}G^2_M}
{1-\frac{t}{4m_N^2}}-2\frac{t}{4m_N^2}G^2_M\tan^2(\theta/2)]
\label{d2}\end{equation}
($\alpha=1/137$, $E$-the incident electron energy)\\
and
\begin{equation}
\sigma_{tot}^{c.m.}(e^+e^-\to N\bar N)=\frac{4\pi\alpha^2\beta_N}{3t}
[|G_M(t)|^2+\frac{2m_N^2}{t}|G_E(t)|^2],\;\;\; \beta_N=\sqrt{1-\frac{4m_N^2}{t}}
\label{d3}
\end{equation}

or
\begin{equation}
\sigma_{tot}^{c.m.}(\bar p p\to e^+e^-)=\frac{2\pi\alpha^2}{3p_{c.m.}\sqrt{t}}
[|G_M(t)|^2+\frac{2m_N^2}{t}|G_E(t)|^2],
\label{d4}\end{equation}
($p_{c.m.}$-antiproton momentum in c.m. system)\\
as there are no interference terms between them.

The isoscalar and isovector Dirac and Pauli ff's are the most suitable for a
construction of various phenomenological models of the nucleon e.m. structure.

In recent years abundant and very accurate data on the nucleon e.m. ff's
appeared. All references concerning the nucleon space-like data
can be found in \ct{1}, besides the recent precise measurements \ct{2-10}.
In the time-like region see \ct{11-19}. Here in particular,
the FENICE experiment in Frascati has measured, besides the proton e.m. ff's
\ct{18}, the magnetic neutron ff in the time-like region \ct{19} for
the first time. There are also valuable results on the magnetic proton
ff \ct{14,15} at higher energies to be measured at FERMILAB.

The present space-like region data will be even considerably improved in the
few GeV region when experiments at TJNAF will be completed.
There are also other experiments under way at MAMI, ELSA, MIT-Bates
involving polarized beams and/or targets in order to give better data in the
space-like region, in particular for the electric ff of the neutron, but also
the magnetic proton and neutron ones.

This all stimulated a new dispersion theoretical analysis \ct{20,21} of the
nucleon e.m. ff data in the space-like region and in the time-like region too
\ct{22}. The latter works are an update and extension of historically the most
competent nucleon ff analysis carried out by H\H{o}hler and collaborators \ct{23}.
However, the model does not allow to describe all the time-like data consistently,
while still giving a good description of the data in the space-like region.

The work presented here was incentivated just by the results in
\ct{20,22} and also by predictions of the spectral function
behaviours in the
framework of the chiral perturbation theory \ct{24}.

In the analysis \ct{20} external constraints on the isovector spectral
functions were used, which consist in
the two-pion continuum effect on the left wing of the
$\rho(770)$-resonance  following  from the $\pi N$-scattering data and pion
e.m. ff behaviour through  the unitarity condition. Further one can see
that our ten-resonance unitary and analytic model, which is just an
improvement of \ct{1,25} and an extension of \ct{26}, contains an explicit
two-pion continuum contribution given by the unitary cut starting
from $t=4m_\pi^2$. Then despite of the fact that here  the unstable $\rho$-meson
is taken into account only as complex conjugate pairs of poles
on the second and third  Riemann sheets of the four
sheeted Riemann
surface, the model itself predicts the strong enhancement of the left
wing of the $\rho(770)$ resonance  in the isovector spectral functions
and moreover, it is  consistent with results of
\ct{20,27}.

Another success of the presented model is the automatic prediction of isoscalar
 nucleon
spectral function behaviours to be consistent with chiral perturbation
theory results \ct{24}. Naturally,  a description of all existing space-like and
time-like nucleon e.m. ff data,  including also FENICE
(Frascati) results from $e^+e^-\to n\bar n$, is achieved for the first time.

The paper is organized as follows. In  section 2. the unitary and analytic
ten-resonance model of the nucleon e.m. structure with canonical
normalizations and asymptotics as predicted by the quark model of hadrons
is constructed. An evaluation of all free parameters of the model
(however, with clear physical meaning) by a fit of all existing data is
carried out in  section 3. In section 4. we predict the isovector and isoscalar
nucleon spectral function behaviours. The last section is devoted to
conclusions and discussion.

\section{Ten resonance unitary and analytic model of nucleon e.m. structure}

The all four sets of nucleon e.m. ff's discussed in the introduction are related
by means of the expressions
\begin{eqnarray}
\nonumber G_E^p(t)&=&G_E^s(t)+G_E^v(t)=F_1^p(t)+\frac{t}{4m^2_p}F_2^p(t)=[F_1^s(t)]+ F_1^v(t)]+\frac{t}{4m_p^2}[F_2^s(t)+F_2^v(t)];\\
 G_M^p(t)&=&G_M^s(t)+G_M^v(t)=F_1^p(t)+F_2^p(t)=[F_1^s(t)+F_1^v(t)]+ [F_2^s(t)+F_2^v(t)]; \label{d5} \\
\nonumber G_E^n(t)&=&G_E^s(t)-G_E^v(t)=F_1^n(t)+\frac{t}{4m_n^2}F_2^n(t)=[F_1^s(t)-F_1^v(t)]+\frac{t}{4m_n^2}[F_2^s(t)-F_2^v(t)];\\
\nonumber G_M^n(t)&=&G^s_M(t)-G_M^v(t)=F_1^n(t)+F_2^n(t)=[F_1^s(t)-F_1^v(t)]+[F_2^s(t)-F_2^v(t)],
\end{eqnarray}
and for the value $t=0$ normalized as follows
\begin{eqnarray}
(i)&\nonumber &
\ G_E^p(0)=1;\ G_M^p(0)=1+\mu_p;\ G_E^n(0)=0;\ G_M^n(0)=\mu_n;\\
(ii)&\nonumber &
\ G_E^s(0)=G_E^v(0)=\frac{1}{2};\ G_M^s(0)=\frac{1}{2}(1+\mu_p+\mu_n);\
 G_M^v(0)=\frac{1}{2}(1+\mu_p-\mu_n);\\
(iii)&\label{d6} &
\ F_1^p(0)=1;\ F_2^p(0)=\mu_p;\ F_1^n(0)=0;\ F_2^n(0)=\mu_n;\\
(iv)&\nonumber&\ F_1^s(0)=F_1^v(0)=\frac{1}{2};\ F_2^s(0)=\frac{1}{2}(\mu_p+\mu_n);\
 F_2^v(0)=\frac{1}{2}(\mu_p-\mu_n),
\end{eqnarray}
where $\mu_p$ and $\mu_n$ are the proton and neutron anomalous magnetic
moments, respectively.

Our ten-resonance unitary and analytic model represents a consistent
unification of the following three fundamental knowledges about the e.m.
ff's:
\begin{itemize}
\item[1.]
The experimental fact of a creation of unstable vector-meson resonances
in the $e^+e^-$-annihilation processes into hadrons.
\item[2.]
The hypothetical analytic properties of the nucleon e.m. ff's.
\item[3.]
The asymptotic behaviour of nucleon e.m. ff's as predicated \ct{28} by the
quark model of hadrons.
\end{itemize}

The most suitable set of ff's for a construction of the model  are the
isoscalar and isovector parts of the Dirac and Pauli ff's to be, in the
first place, saturated by the isoscalar and isovector vector mesons
possessing the quantum numbers of the photon. Here we stand up for a view
that as there are no data on the nucleon e.m. ff's at the region
$0<t<4m_N^2$ of a manifestation of the majority of resonances  under
consideration, the resonance parameters have to be always fixed at the
world averaged values and then investigated their consistency with existing
ff data in other regions and also with principles on the base of which the
considered model is constructed.

In Review of Particle Physics \ct{29} we find just 5 isoscalar resonances
$\omega$(782), $\phi$(1020), $\omega^,$(1420), $\omega^{,,}$(1600),
$\phi^,$(1680) with required properties. However, one finds only 3
isovector resonances $\rho$(770), $\rho^,$(1450), $\rho^{,,}$(1700) with
quantum numbers of the photon there. On the other hand, we have obtained an
experience in \ct{1,25,26,30} that the most stable description of existing
data is obtained if equal number of isoscalar and isovector resonances in
the investigated model is taken into account. Therefore in the isovector
Dirac and Pauli ff's we consider also the third excited state of the
$\rho$-meson, $\rho^{,,,}$(2150), revealed in \ct{31}, and moreover, we also
introduce hypothetically the fourth excited state of the
$\rho$-meson $\rho^{,,,,}$(?), the mass and width of which  are
free parameters of the model. As one can see further in a comparison of
the model with all existing data, those resonance parameters will be found
to be quite reasonable and thus they provide simultaneous perfect description of the
space-like and time-like nucleon ff data, including also the FENICE
(Frascati) results  on the neutron.

Now, in order to take into account
the experimental fact of a creation of vector-meson resonances in
$e^+e^-$ annihilation into hadrons, we start with the
vector-meson-dominance (VMD) parametrization of the isoscalar and isovector
parts of the Dirac and Pauli ff's
\begin{eqnarray}
\nonumber F_1^s(t)=\sum_{\sa,\sd,\sbb,\sc,\se}\mmm s\fff 1{s};
& &F_1^v(t)=\sum_{\va,\vb,\vc,\vd,\ve}\mmm v\fff 1{v};\\
F_2^s(t)=\sum_{\sa,\sd,\sbb,\sc,\se}\mmm s\fff 2{s};
& &F_2^v(t)=\sum_{\va,\vb,\vc,\vd,\ve}\mmm v\fff 2{v},\label{d7}
\end{eqnarray}
where $m_s$ and $m_v$ are isoscalar and isovector vector-meson
masses,  $f^{(1)}_{sNN}$, $f^{(1)}_{vNN}$ and $f^{(2)}_{sNN}$,
$f^{(2)}_{vNN}$ are vector and tensor vector-meson-nucleon coupling
constants and $f_s$, $f_v$ are the universal vector-meson coupling
constants to be determined in a vector-meson decay into two charged
leptons. Here in the isoscalar ff's  also $\phi$(1020) and
$\phi^,$(1680) meson contributions are considered as there are clear
indications on the strange quark content in the nucleon and  the OZI
rule violation as well.

The expressions (\ref{d7}) do not govern neither the normalization conditions
(\ref{d6}), nor the asymptomatic behaviour
\begin{eqnarray}
t^{i+1}F_i^{s,v}(t)_{|t|\to\infty}\sim {constant},\;\;\;\;i=1,2
\label{d8}
\end{eqnarray}
to be consistent up to logarithmic correction with results as predicted
\ct{28} by the quark model of hadrons. However, any serious attempt to
describe the present experimental data on the nucleon e.m. ff's has to
account for these constraints. Their explicit requirement in (\ref{d7}) leads to
four systems of algebraic equations

\begin{eqnarray}
\nonumber {\rm I}. & &\sum_{\sa,\sd,\sbb,\sc,\se}\fff 1{s}=\frac{1}{2}\\
\  & &\sum_{\sa,\sd,\sbb,\sc,\se}\fff 1{s}m_s^2=0\label{d9}\\
\nonumber\ & &\\
\nonumber {\rm II}. & &\sum_{\va,\vb,\vc,\vd,\ve}\fff 1{v}=\frac{1}{2}\\
\  & &\sum_{\va,\vb,\vc,\vd,\ve}\fff 1{v}m_v^2=0\label{d10}\\
\nonumber\ & &\\
\nonumber {\rm III}. & &\sum_{\sa,\sd,\sbb,\sc,\se}\fff 2{s}=\frac{1}{2}(\mu_p+\mu_n)\\
\   & &\sum_{\sa,\sd,\sbb,\sc,\se}\fff 2{s}m_s^2=0 \label{d11}\\
\nonumber & &\fff 2{\sa}\mcsa(\mcsd+\mcsb+\mcsc+\mcse)+\\
\nonumber
& &\fff 2{\sd}\mcsd(\mcsa+\mcsb+\mcsc+\mcse)+\\
\nonumber
& &\fff 2{\sbb}\mcsb(\mcsd+\mcsa+\mcsc+\mcse)+\\
\nonumber
& &\fff 2{\sc}\mcsc(\mcsd+\mcsb+\mcsa+\mcse)+\\
\nonumber
& &\fff 2{\se}\mcse(\mcsd+\mcsb+\mcsc+\mcsa)=0\\
\nonumber\ & &\\
\nonumber {\rm IV}. & &\sum_{\va,\vb,\vc,\vd,\ve}\fff 2{v}=\frac{1}{2}(\mu_p-\mu_n)\\
& &\sum_{\va,\vb,\vc,\vd,\ve}\fff 2{v}m_v^2=0 \label{d12}\\
\nonumber
& &\fff 2{\va}\mcva(\mcvb+\mcvc+\mcvd+\mcve)+\\
\nonumber
& &\fff 2{\vb}\mcvb(\mcva+\mcvc+\mcvd+\mcve)+\\
\nonumber
& &\fff 2{\vc}\mcvc(\mcvb+\mcva+\mcvd+\mcve)+\\
\nonumber
& &\fff 2{\vd}\mcvd(\mcvb+\mcvc+\mcva+\mcve)+\\
\nonumber
& &\fff 2{\ve}\mcve(\mcvb+\mcvc+\mcvd+\mcva)=0
\end{eqnarray}

for $\fff 1{s}$, $\fff 1{v}$, $\fff 2{s}$, and $\fff 2{v}$, which reduce a
number of free parameters of the constructed model remarkably.

Solutions of the (\ref{d9})-(\ref{d12})can be chosen in the following form
\begin{eqnarray}
\nonumber {\rm I.} \;\;\; \fff {1}{\sbb}&=&\frac{1}{2}\frac{\mcsc}{\mcsc-\mcsb}
  -\fff {1}{\sa}\zll{\mcsc}{\mcsa}{\mcsc}{\mcsb}-\\
&-&\fff {1}\sd\zll{\mcsc}{\mcsd}{\mcsc}{\mcsb}+
  \fff {1}\se\zll{\mcse}{\mcsc}{\mcsc}{\mcsb}\label{d13}\\
\nonumber
\fff {1}{\sc}&=&-\frac{1}{2}\frac{\mcsb}{\mcsc-\mcsb}
  +\fff {1}{\sa}\zll{\mcsb}{\mcsa}{\mcsc}{\mcsb}+\\
\nonumber
&+&\fff {1}\sd\zll{\mcsb}{\mcsd}{\mcsc}{\mcsb}-
  \fff {1}\se\zll{\mcse}{\mcsb}{\mcsc}{\mcsb}\\
\nonumber
& &\qquad\\
\nonumber
{\rm II.} \;\;\;\fff {1}{\vb}&=&\frac{1}{2}\frac{\mcvc}{\mcvc-\mcvb}
  -\fff {1}{\va}\zll{\mcvc}{\mcva}{\mcvc}{\mcvb}+\\
&+&\fff {1}\vd\zll{\mcvd}{\mcvc}{\mcvc}{\mcvb}
  +\fff {1}\ve\zll{\mcve}{\mcvc}{\mcvc}{\mcvb}\label{d14}\\
\nonumber
 \fff {1}{\vc}&=&-\frac{1}{2}\frac{\mcvb}{\mcvc-\mcvb}
  +\fff {1}{\va}\zll{\mcvb}{\mcva}{\mcvc}{\mcvb}\\
\nonumber
&-&\fff {1}\vd\zll{\mcvd}{\mcvb}{\mcvc}{\mcvb}
  -\fff {1}\ve\zll{\mcve}{\mcvb}{\mcvc}{\mcvb}\\
\nonumber
& &\qquad\\
\nonumber
{\rm III.}\;\;\; \fff 2{\sa}&=&\frac{1}{2}(\mu_p+\mu_n)\zzl\mcsc\mcsb\mcsc\mcsa\mcsb\mcsa-\\
\nonumber
  &-&\fff 2{\sd}\zlll\mcsc\mcsd\mcsb\mcsd\mcsc\mcsa\mcsb\mcsa-\\
\nonumber
  &-&\fff 2{\se}\zlll\mcse\mcsc\mcse\mcsb\mcsc\mcsa\mcsb\mcsa\\
\nonumber
 \fff 2{\sbb}&=&-\frac{1}{2}(\mu_p+\mu_n)\zzl\mcsc\mcsa\mcsc\mcsb\mcsb\mcsa-\\
 &-&\fff 2{\sd}\zlll\mcsc\mcsd\mcsd\mcsa\mcsc\mcsb\mcsb\mcsa+\label{d15}\\
\nonumber
  &+&\fff 2{\se}\zlll\mcse\mcsc\mcse\mcsa\mcsc\mcsb\mcsb\mcsa\\
\nonumber
 \fff 2{\sc}&=&\frac{1}{2}(\mu_p+\mu_n)\zzl\mcsb\mcsa\mcsc\mcsb\mcsc\mcsa+\\
\nonumber
 &+&\fff 2{\sd}\zlll\mcsb\mcsd\mcsd\mcsa\mcsc\mcsb\mcsc\mcsa-\\
\nonumber
 &-&\fff 2{\se}\zlll\mcse\mcsb\mcse\mcsa\mcsc\mcsb\mcsc\mcsa\\
\nonumber
& &\qquad\\
\nonumber
{\rm IV.}\;\;\; \fff 2{\va}&=&\frac{1}{2}(\mu_p-\mu_n)\zzl\mcvc\mcvb\mcvc\mcva\mcvb\mcva-\\
\nonumber
   &-&\fff 2{\vd}\zlll\mcvd\mcvc\mcvd\mcvb\mcvc\mcva\mcvb\mcva-\\
\nonumber
   &-&\fff 2{\ve}\zlll\mcve\mcvc\mcve\mcvb\mcvc\mcva\mcvb\mcva\\
\nonumber
 \fff 2{\vb}&=&-\frac{1}{2}(\mu_p-\mu_n)\zzl\mcvc\mcva\mcvc\mcvb\mcvb\mcva+\\
   &+&\fff 2{\vd}\zlll\mcvd\mcvc\mcvd\mcva\mcvc\mcvb\mcvb\mcva+\label{d16}\\
\nonumber
   &+&\fff 2{\ve}\zlll\mcve\mcvc\mcve\mcva\mcvc\mcvb\mcvb\mcva\\
\nonumber
 \fff 2{\vc}&=&\frac{1}{2}(\mu_p-\mu_n)\zzl\mcvb\mcva\mcvc\mcvb\mcvc\mcva-\\
\nonumber
   &-&\fff 2{\vd}\zlll\mcvd\mcvb\mcvd\mcva\mcvc\mcvb\mcvc\mcva-\\
\nonumber
   &-&\fff 2{\ve}\zlll\mcve\mcvb\mcve\mcva\mcvc\mcvb\mcvc\mcva,
\end{eqnarray}
which transform the original parametrizations (\ref{d7}) of the isoscalar and
isovector Dirac and Pauli nucleon ff's still into the zero-width VMD
expressions
\begin{eqnarray}
\nonumber F^s_1(t)&=&\frac{1}{2}\mmt\sc\sbb+\\
  \nonumber &+&\left\{\mmt\sc\sa\zll\mcsc\mcsa\mcsc\mcsb-\mmt\sbb\sa\zll\mcsb\mcsa\mcsc\mcsb-\right.\\
  &-&\left.\mmt\sc\sbb\right\}\fff 1\sa+\label{d17}\\
  \nonumber &+&\left\{\mmt\sc\sd\zll\mcsc\mcsd\mcsc\mcsb-\mmt\sbb\sd\zll\mcsb\mcsd\mcsc\mcsb-\right.\\
  \nonumber &-&\left.\mmt\sc\sbb\right\}\fff 1\sd-\\
  \nonumber &-&\left\{\mmt\se\sc\zll\mcse\mcsc\mcsc\mcsb-\mmt\se\sbb\zll\mcse\mcsb\mcsc\mcsb+\right.\\
  \nonumber &+&\left.\mmt\sc\sbb\right\}\fff 1\se,\\
\nonumber & &\qquad\\
\nonumber F^v_1(t)&=&\frac{1}{2}\mmt\vc\vb+\\
   \nonumber &+&\left\{\mmt\vc\va\zll\mcvc\mcva\mcvc\mcvb-\mmt\vb\va\zll\mcvb\mcva\mcvc\mcvb-\right.\\
   &-&\left.\mmt\vc\vb\right\}\fff 1\va+\label{d18}\\
   \nonumber &+&\left\{\mmt\vd\vb\zll\mcvd\mcvb\mcvc\mcvb-\mmt\vd\vc\zll\mcvd\mcvc\mcvc\mcvb-\right.\\
   \nonumber &-&\left.\mmt\vc\vb\right\}\fff 1\vd-\\
   \nonumber &-&\left\{\mmt\ve\vc\zll\mcve\mcvc\mcvc\mcvb-\mmt\ve\vb\zll\mcve\mcvb\mcvc\mcvb+\right.\\
   \nonumber &+&\left.\mmt\vc\vb\right\}\fff 1\ve,\\
\nonumber & &\qquad\\
\nonumber F_2^s(t)&=&\frac{1}{2}(\mu_p+\mu_n)\mtt\sc\sbb\sa+\\
   \nonumber &+&\left\{\mtt\sc\sd\sa\zlll\mcsc\mcsd\mcsd\mcsa\mcsc\mcsb\mcsb\mcsa+\right.\\
   \nonumber &+&\mtt\sc\sbb\sd\zlll\mcsc\mcsd\mcsb\mcsd\mcsc\mcsa\mcsb\mcsa-\\
   \nonumber &-&\mtt\sbb\sd\sa\zlll\mcsb\mcsd\mcsd\mcsa\mcsc\mcsb\mcsc\mcsa-\\
   &-&\left.\mtt\sc\sbb\sa\right\}\fff 2\sd+\label{d19}\\
   \nonumber &+&\left\{\mtt\se\sc\sbb\zlll\mcse\mcsc\mcse\mcsb\mcsc\mcsa\mcsb\mcsa-\right.\\
   \nonumber &-&\mtt\se\sc\sa\zlll\mcse\mcsc\mcse\mcsa\mcsc\mcsb\mcsb\mcsa+\\
   \nonumber &+&\mtt\se\sbb\sa\zlll\mcse\mcsb\mcse\mcsa\mcsc\mcsb\mcsc\mcsa-\\
   \nonumber &-&\left.\mtt\sc\sbb\sa\right\}\fff 2\se,\\
\nonumber & &\qquad\\
\nonumber F_2^v(t)&=&\frac{1}{2}(\mu_p-\mu_n)\mtt\vc\vb\va+\\
  \nonumber &+&\left\{\mtt\vd\vb\va\zlll\mcvd\mcvb\mcvd\mcva\mcvc\mcvb\mcvc\mcva-\right.\\
  \nonumber &-&\mtt\vd\vc\va\zlll\mcvd\mcvc\mcvd\mcva\mcvc\mcvb\mcvb\mcva+\\
  \nonumber &+&\mtt\vd\vc\vb\zlll\mcvd\mcvc\mcvd\mcvb\mcvc\mcva\mcvb\mcva-\\
  &-&\left.\mtt\vc\vb\va\right\}\fff 2\vd+\label{d20}\\
  \nonumber &+&\left\{\mtt\ve\vb\va\zlll\mcve\mcvb\mcve\mcva\mcvc\mcvb\mcvc\mcva-\right.\\
  \nonumber &-&\mtt\ve\vc\va\zlll\mcve\mcvc\mcve\mcva\mcvc\mcvb\mcvb\mcva+\\
  \nonumber &+&\mtt\ve\vc\vb\zlll\mcve\mcvc\mcve\mcvb\mcvc\mcva\mcvb\mcva-\\
  \nonumber &-&\left.\mtt\vc\vb\va\right\}\fff 2\ve,
\end{eqnarray}
however, these are already automatically normalized and they govern the
required asymptotic behaviour (\ref{d8}).
Despite the latter properties the model
is unable to reproduce the existing experimental information properly and
only its unitarization leads to a correct simultaneous description of the
space-like and time-like data.

It is well known that the unitarity condition requires the imaginary part of
the nucleon e.m. ff's to be different from zero  above the lowest branch
point $t_0$ and, moreover, it just predicts its smoothly varying behaviour
(see e.g. \ct{20}, \ct{27}).

Further we show that the unitarization of the model (\ref{d17})-(\ref{d20}) can be
achieved by application of the following special non-linear transformations
\begin{eqnarray}
\nonumber t=t_0^s-\frac{4(t_{in}^{1s}-t_0^s)}{[1/V-V]^2},
& &t=t_0^s-\frac{4(t_{in}^{2s}-t_0^s)}{[1/U-U]^2},\\
t=t_0^v-\frac{4(t_{in}^{1v}-t_0^v)}{[1/W-W]^2},
& &t=t_0^v-\frac{4(t_{in}^{2v}-t_0^v)}{[1/X-X]^2},\label{d21}
\end{eqnarray}
respectively, and a subsequent incorporation of the non-zero values of
vector meson widths. There are also another expressions utilized for
the vector meson masses squared
\begin{eqnarray}
\nonumber\mc s=t_0^s-\frac{4(t_{in}^{1s}-t_0^s)}{[1/V_{s0}-V_{s0}]^2},
& &\mc s=t_0^s-\frac{4(t_{in}^{2s}-t_0^s)}{[1/U_{s0}-U_{s0}]^2},\\
\mc v=t_0^v-\frac{4(t_{in}^{1v}-t_0^v)}{[1/W_{v0}-W_{v0}]^2},
& &\mc v=t_0^v-\frac{4(t_{in}^{2v}-t_0^v)}{[1/X_{v0}-X_{v0}]^2},\label{d22}
\end{eqnarray}
and identities
\begin{eqnarray}
\nonumber 0=t_0^s-\frac{4(t_{in}^{1s}-t_0^s)}{[1/V_N-V_N]^2},
& &0=t_0^s-\frac{4(t_{in}^{2s}-t_0^s)}{[1/U_N-U_N]^2},\\
0=t_0^v-\frac{4(t_{in}^{1v}-t_0^v)}{[1/W_N-W_N]^2},
& &0=t_0^v-\frac{4(t_{in}^{2v}-t_0^v)}{[1/X_N-X_N]^2},\label{d23}
\end{eqnarray}
following from (\ref{d21}), where $V_{s0}$, $W_{v0}$, $U_{s0}$, $X_{v0}$ are
the zero-width (therefore they have a subindex 0) VMD poles and $V_N$, $W_N$,
$U_N$, $X_N$ are the normalization points (corresponding to $t=0$) in the
$V$, $W$, $U$, $X$ planes, respectively, and $t_0^s=9m_\pi^2$, $t_0^v=4m_\pi^2$,
$t_{in}^{1s}$, $t_{in}^{2s}$, $t_{in}^{1v}$, $t_{in}^{2v}$ are square-root
branch points as it is transparent from the inverse transformations to (\ref{d21}), e.g.
\begin{equation}
V(t)=i\frac
{\sqrt{\left (\frac{t_{in}^{1s}-t_0^s}{t_0^s}\right )^{1/2}+\left (\frac{t-t_0^s}{t_0^s}\right )^{1/2}}-
 \sqrt{\left (\frac{t_{in}^{1s}-t_0^s}{t_0^s}\right )^{1/2}-\left (\frac{t-t_0^s}{t_0^s}\right )^{1/2}}}
{\sqrt{\left (\frac{t_{in}^{1s}-t_0^s}{t_0^s}\right )^{1/2}+\left (\frac{t-t_0^s}{t_0^s}\right )^{1/2}}+
 \sqrt{\left (\frac{t_{in}^{1s}-t_0^s}{t_0^s}\right )^{1/2}-\left (\frac{t-t_0^s}{t_0^s}\right )^{1/2}}}
\label{d24}\end{equation}
and similarly for $W(t)$, $U(t)$ and  $X(t)$.

Really, the relations (\ref{d21})-(\ref{d23}) first transform every t-dependent term and
every constant term consisting of a ratio of mass differences in
(\ref{d17})-(\ref{d20})
into a new form as follows. For instance the term $\mc\sa/(\mc\sa-t)$
in (\ref{d17}) is transformed into the following factorized form:
\begin{equation}
\mmm\omega=\frac{\mcsa-0}{\mcsa-t}=\left(\frac{1-V^2}{1-V_N^2}\right)^2\frac
 {(V_N-V_{\sam})(V_N+V_{\sam})(V_N-1/V_{\sam})(V_N+1/V_{\sam})}
 {(V-V_{\sam})(V+V_{\sam})(V-1/V_{\sam})(V+1/V_{\sam})}.\label{d25}
\end{equation}
The constant mass terms, e.g. $(\mcsb-\mcsa)/(\mcsc-\mcsb)$ also from (\ref{d17}),
become as follows
\begin{eqnarray}
& &\frac{\mcsb-\mcsa}{\mcsc-\mcsb}=
\frac{(\mcsb-0)-(\mcsa-0)}{(\mcsc-0)-(\mcsb-0)}=\label{d26}\\
\nonumber & &=\left [\frac{(V_N-V_{\sbbm})(V_N+V_{\sbbm})(V_N-1/V_{\sbbm})(V_N+1/V_{\sbbm})}
  {(V_{\sbbm}-1/V_{\sbbm})^2}\right.-\\
\nonumber & &\left.
-\frac{(V_N-V_{\sam})(V_N+V_{\sam})(V_N-1/V_{\sam})(V_N+1/V_{\sam})}
  {(V_{\sam}-1/V_{\sam})^2}\right ]/\\
\nonumber & &\left[\frac{(V_N-V_{\scm})(V_N+V_{\scm})(V_N-1/V_{\scm})(V_N+1/V_{\scm})}
  {(V_{\scm}-1/V_{\scm})^2}\right.-\\
\nonumber & &\left.-\frac{(V_N-V_{\sbbm})(V_N+V_{\sbbm})(V_N-1/V_{\sbbm})(V_N+1/V_{\sbbm})}
  {(V_{\sbbm}-1/V_{\sbbm})^2}\right ]=\\
\nonumber & &=
\frac{C_{{\omega'}_0}^{1s}-C_{{\omega}_0}^{1s}}{C_{{\omega''}_0}^{1s}-C_{{\omega'}_0}^{1s}}.
\end{eqnarray}

Then by utilization of the relations between complex and complex conjugate
values of the
corresponding zero-width VMD pole positions in the $V$, $W$, $U$, $X$ planes
\begin{eqnarray}
\nonumber & &V_{\sam}=-V_{\sam}^*;\;V_{\sdm}=-V_{\sdm}^*;\;V_{\sbbm}=-V_{\sbbm}^*;\;
 V_{\scm}=1/V_{\scm}^*;\;V_{\sem}=1/V_{\sem}^*\\
& &W_{\vam}=-W_{\vam}^*;\;W_{\vbm}=-W_{\vbm}^*;\;W_{\vcm}=-W_{\vcm}^*;\;
 W_{\vdm}=1/W_{\vdm}^*;\;W_{\vem}=1/W_{\vem}^*\label{d27}\\
\nonumber & &U_{\sam}=-U_{\sam}^*;\;U_{\sdm}=-U_{\sdm}^*;\;U_{\sbbm}=-U_{\sbbm}^*;\;
 U_{\scm}=1/U_{\scm}^*;\;U_{\sem}=1/U_{\sem}^*\\
\nonumber & &X_{\vam}=-X_{\vam}^*;\;X_{\vbm}=-X_{\vbm}^*;\;X_{\vcm}=-X_{\vcm}^*;\;
 X_{\vdm}=1/X_{\vdm}^*;\;X_{\vem}=1/X_{\vem}^*
\end{eqnarray}
following from the fact that in a fitting procedure we find
\begin{eqnarray}
\nonumber
& &\mcsa-\Gamma^2_{\sa}/4<t_{in}^{1s};\;
\mcsd-\Gamma^2_{\sd}/4<t_{in}^{1s};\;
\mcsb-\Gamma^2_{\sbb}/4<t_{in}^{1s};\;\\
\nonumber
& &\mcsc-\Gamma^2_{\sc}/4>t_{in}^{1s};\;
\mcse-\Gamma^2_{\se}/4>t_{in}^{1s};\\
\nonumber
& &\mcsa-\Gamma^2_{\sa}/4<t_{in}^{2s};\;
\mcsd-\Gamma^2_{\sd}/4<t_{in}^{2s};\;
\mcsb-\Gamma^2_{\sbb}/4<t_{in}^{2s};\;\\
& &\mcsc-\Gamma^2_{\sc}/4>t_{in}^{2s};\;
\mcse-\Gamma^2_{\se}/4>t_{in}^{2s};\label{d28}\\
\nonumber
& &\mcva-\Gamma^2_{\va}/4<t_{in}^{1v};\;
\mcvb-\Gamma^2_{\vb}/4<t_{in}^{1v};\;
\mcvc-\Gamma^2_{\vc}/4<t_{in}^{1v};\;\\
\nonumber
& &\mcvd-\Gamma^2_{\vd}/4>t_{in}^{1v};\;
\mcve-\Gamma^2_{\ve}/4>t_{in}^{1v};\\
\nonumber
& &\mcva-\Gamma^2_{\va}/4<t_{in}^{2v};\;
\mcvb-\Gamma^2_{\vb}/4<t_{in}^{2v};\;
\mcvc-\Gamma^2_{\vc}/4<t_{in}^{2v};\;\\
\nonumber
& &\mcvd-\Gamma^2_{\vd}/4>t_{in}^{2v};\;
\mcve-\Gamma^2_{\ve}/4>t_{in}^{2v};
\end{eqnarray}
and subsequent introduction of the non-zero values of vector-meson widths
$\Gamma\not= 0$ by the substitutions
\begin{equation}
\mc s\to(m_s-i\frac{\Gamma_s}{2})^2;\;\;\;\;\mc v\to(m_v-i\frac{\Gamma_v}{2})^2,
\label{d29}\end{equation}
one gets, for every isoscalar and isovector Dirac and Pauli ff, one analytic
function in the whole complex $t$-plane besides two right-hand cuts of the
following forms
\begin{eqnarray}
\nonumber F^s_1[V(t)]&=&\left(\frac{1-V^2}{1-V_N^2}\right)^4\left\{\frac{1}{2}\eee V\sc.\right.\\
\nonumber &.&\xxx V\sbb+\\
\nonumber &+&\left[\eee V\sc.\right.\\
\nonumber &.&\xxx V\sa.\ccc {1s}\sc\sa\sc\sbb-\\
\nonumber &-&\xxx V\sbb.\\
\nonumber &.&\xxx V\sa.\ccc {1s}\sbb\sa\sc\sbb-\\
\nonumber &-&\eee V\sc.\\
\nonumber &.&\left.\xxx V\sbb\right]\fff 1\sa+\\
\nonumber &+&\left[\eee V\sc.\right.\\
\nonumber &.&\xxx V\sd.\ccc {1s}\sc\sd\sc\sbb-\\
          &-&\xxx V\sbb.\label{d30}\\
\nonumber &.&\xxx V\sd.\ccc {1s}\sbb\sd\sc\sbb-\\
\nonumber &-&\eee V\sc.\\
\nonumber &.&\left.\xxx V\sbb\right]\fff 1\sd-\\
\nonumber &-&\left[\eee V\se.\right.\\
\nonumber &.&\eee V\sc.\ccc {1s}\se\sc\sc\sbb-\\
\nonumber &-&\eee V\se.\\
\nonumber &.&\xxx V\sbb.\ccc {1s}\se\sbb\sc\sbb+\\
\nonumber &+&\eee V\sc.\\
\nonumber &.&\left.\left.\xxx V\sbb\right]\fff 1\se\right\}\\
\nonumber \;\\
\nonumber F^v_1[W(t)]&=&\left(\frac{1-W^2}{1-W_N^2}\right)^4\left\{\frac{1}{2}\xxx W\vc.\right.\\
 \nonumber &.&\xxx W\vb+\\
 \nonumber &+&\left[\xxx W\vc.\right.\\
 \nonumber &.&\xxx W\va.\ccc {1v}\vc\va\vc\vb-\\
 \nonumber &-&\xxx W\vb.\\
 \nonumber &.&\xxx W\va.\ccc {1v}\vb\va\vc\vb-\\
 \nonumber &-&\xxx W\vc.\\
 \nonumber &.&\left.\xxx W\vb\right]\fff 1\va+\\
 \nonumber &+&\left[\eee W\vd.\right.\\
 \nonumber &.&\xxx W\vb.\ccc {1v}\vd\vb\vc\vb-\\
           &-&\eee W\vd.\label{d31}\\
 \nonumber &.&\xxx W\vc.\ccc {1v}\vd\vc\vc\vb-\\
 \nonumber &-&\xxx W\vc.\\
 \nonumber &.&\left.\xxx W\vb\right]\fff 1\vd-\\
 \nonumber &-&\left[\eee W\ve.\right.\\
 \nonumber &.&\xxx W\vc.\ccc {1v}\ve\vc\vc\vb-\\
 \nonumber &-&\eee W\ve.\\
 \nonumber &.&\xxx W\vb.\ccc {1v}\ve\vb\vc\vb+\\
 \nonumber &+&\xxx W\vc.\\
 \nonumber &.&\left.\left.\xxx W\vb\right]\fff 1\ve\right\}
\end{eqnarray}

\begin{eqnarray}
\nonumber F_2^s[U(t)]&=&\left(\frac{1-U^2}{1-U_N^2}\right)^6\left\{\frac{1}{2}(\mu_p+\mu_n)\eee U\sc.\right.\\
 \nonumber &.&\xxx U\sbb.\\
 \nonumber &.&\xxx U\sa+\\
 \nonumber &+&\left[\eee U\sc.\right.\\
 \nonumber &.&\xxx U\sd.\\
 \nonumber &.&\xxx U\sa.\ccc{2s}\sc\sd\sc\sbb.\ccc {2s}\sd\sa\sbb\sa+\\
 \nonumber &+&\eee U\sc.\\
 \nonumber &.&\xxx U\sbb.\\
 \nonumber &.&\xxx U\sd.\ccc{2s}\sc\sd\sc\sa.\ccc {2s}\sbb\sd\sbb\sa-\\
 \nonumber &-&\xxx U\sbb.\\
 \nonumber &.&\xxx U\sd.\\
 \nonumber &.&\xxx U\sa.\ccc{2s}\sbb\sd\sc\sbb.\ccc {2s}\sd\sa\sc\sa-\\
           &-&\eee U\sc.\label{d32}\\
 \nonumber &.&\xxx U\sbb.\\
 \nonumber &.&\left.\xxx U\sa\right]\fff 2\sd+\\
 \nonumber &+&\left[\eee U\se.\right.\\
 \nonumber &.&\eee U\sc.\\
 \nonumber &.&\xxx U\sbb.\ccc{2s}\se\sc\sc\sa.\ccc {2s}\se\sbb\sbb\sa-\\
 \nonumber &-&\eee U\se.\\
 \nonumber &.&\eee U\sc.\\
 \nonumber &.&\xxx U\sa.\ccc{2s}\se\sc\sc\sbb.\ccc {2s}\se\sa\sbb\sa+\\
 \nonumber &+&\eee U\se.\\
 \nonumber &.&\xxx U\sbb.\\
 \nonumber &.&\xxx U\sa.\ccc{2s}\se\sbb\sc\sbb.\ccc {2s}\se\sa\sc\sa-\\
 \nonumber &-&\eee U\sc.\\
 \nonumber &.&\xxx U\sbb.\\
 \nonumber &.&\left.\left.\xxx U\sa\right]\fff 2\se\right\}
\end{eqnarray}

\begin{eqnarray}
\nonumber F_2^v[X(t)]&=&\left(\frac{1-X^2}{1-X_N^2}\right)^6\left\{\frac{1}{2}(\mu_p-\mu_n)\xxx X\vc.\right.\\
 \nonumber &.&\xxx X\vb.\\
 \nonumber &.&\xxx X\va+\\
 \nonumber &+&\left[\eee X\vd.\right.\\
 \nonumber &.&\xxx X\vb.\\
 \nonumber &.&\xxx X\va.\ccc{2v}\vd\vb\vc\vb.\ccc {2v}\vd\va\vc\va-\\
 \nonumber &-&\eee X\vd.\\
 \nonumber &.&\xxx X\vc.\\
 \nonumber &.&\xxx X\va.\ccc{2v}\vd\vc\vc\vb.\ccc {2v}\vd\va\vb\va+\\
 \nonumber &+&\eee X\vd.\\
 \nonumber &.&\xxx X\vc.\\
 \nonumber &.&\xxx X\vb.\ccc{2v}\vd\vc\vc\va.\ccc {2v}\vd\vb\vb\va-\\
 \nonumber &-&\xxx X\vc.\\
           &.&\xxx X\vb.\label{d33}\\
 \nonumber &.&\left.\xxx X\va\right]\fff 2\vd+\\
 \nonumber &+&\left[\eee X\ve.\right.\\
 \nonumber &.&\xxx X\vb.\\
 \nonumber &.&\xxx X\va.\ccc{2v}\ve\vb\vc\vb.\ccc {2v}\ve\va\vc\va-\\
 \nonumber &-&\eee X\ve.\\
 \nonumber &.&\xxx X\vc.\\
 \nonumber &.&\xxx X\va.\ccc{2v}\ve\vc\vc\vb.\ccc {2v}\ve\va\vb\va+\\
 \nonumber &+&\eee X\ve.\\
 \nonumber &.&\xxx X\vc.\\
 \nonumber &.&\xxx X\vb.\ccc{2v}\ve\vc\vc\va.\ccc {2v}\ve\vb\vb\va-\\
 \nonumber &-&\xxx X\vc.\\
 \nonumber &.&\xxx X\vb.\\
 \nonumber &.&\left.\left.\xxx X\va\right]\fff 2\ve\right\}
\end{eqnarray}
where
\begin{eqnarray}
\nonumber & &C_r^{1s}=\frac{(V_N-V_r)(V_N-V_r^*)(V_N-1/V_r)(V_N-1/V_r^*)}
               {-(V_r-1/V_r)(V_r^*-1/V_r^*)}; \qquad r=\omega,\phi,\omega^,\\
& &C_l^{1s}=\frac{(V_N-V_l)(V_N-V_l^*)(V_N+V_l)(V_N+V_l^*)}
               {-(V_l-1/V_l)(V_l^*-1/V_l^*)}; \qquad l=\omega^{,,},\phi^,\label{d34}\\
\nonumber \;\\
\nonumber & &C_k^{1v}=\frac{(W_N-W_k)(W_N-W_k^*)(W_N-1/W_k)(W_N-1/W_k^*)}
               {-(W_k-1/W_k)(W_k^*-1/W_k^*)}; \qquad k=\rho,\rho^,,\rho^{,,},\\
& &C_n^{1v}=\frac{(W_N-W_n)(W_N-W_n^*)(W_N+W_n)(W_N+W_n^*)}
               {-(W_n-1/W_n)(W_n^*-1/W_n^*)}; \qquad n=\rho^{,,,},\rho^{,,,,}\label{d35}\\
\nonumber \;\\
\nonumber & &C_r^{2s}=\frac{(U_N-U_r)(U_N-U_r^*)(U_N-1/U_r)(U_N-1/U_r^*)}
               {-(U_r-1/U_r)(U_r^*-1/U_r^*)}; \qquad r=\omega,\phi,\omega^,\\
& &C_l^{2s}=\frac{(U_N-U_l)(U_N-U_l^*)(U_N+U_l)(U_N+U_l^*)}
               {-(U_l-1/U_l)(U_l^*-1/U_l^*)}; \qquad l=\omega^{,,},\phi^,\label{d36}\\
\nonumber \;\\
\nonumber & &C_k^{2v}=\frac{(X_N-X_k)(X_N-X_k^*)(X_N-1/X_k)(X_N-1/X_k^*)}
               {-(X_k-1/X_k)(X_k^*-1/X_k^*)}; \qquad k=\rho,\rho^,,\rho^{,,},\\
& &C_n^{2v}=\frac{(X_N-X_n)(X_N-X_n^*)(X_N+X_n)(X_N+X_n^*)}
               {-(X_n-1/X_n)(X_n^*-1/X_n^*)}; \qquad n=\rho^{,,,},\rho^{,,,,}\label{d37}
\end{eqnarray}

As a result each ff is defined on a four-sheeted Riemann surface in $t$-variable
with poles corresponding to vector-meson resonances placed on unphysical
sheets.

The expressions (\ref{d30})-(\ref{d33}), together with the relations
(\ref{d5}), represent just a
ten-resonance unitary and analytic model of the nucleon e.m. structure with
canonical normalizations (\ref{d6}) and the correct asymptotic behaviours
as predicted
by the quark model of hadrons. In the next sections this model is used to
analyze  all existing nucleon e.m. ff data and to obtain subsequent predictions.

\section{Analysis of all existing space-like and time-like data}

Our ten-resonance unitary and analytic model of the nucleon e.m. structure
depends after all on the following parameters
$t_{in}^{1s}$, $t_{in}^{1v}$, $t_{in}^{2s}$, $t_{in}^{2v}$,
$m_\omega$, $\Gamma_\omega$, $m_\phi$, $\Gamma_\phi$,
$m_{\omega^,}$, $\Gamma_{\omega^,}$, $m_{\omega^{,,}}$, $\Gamma_{\omega^{,,}}$,
$m_{\phi^{,}}$, $\Gamma_{\phi^,}$,
$m_\rho$, $\Gamma_\rho$, $m_{\rho^,}$, $\Gamma_{\rho^,}$, $m_{\rho^{,,}}$,
$\Gamma_{\rho^{,,}}$,
$m_{\rho^{,,,}}$, $\Gamma_{\rho^{,,,}}$, $m_{\rho^{,,,,}}$,
$\Gamma_{\rho^{,,,,}}$,
$f_{\omega NN}^{(1)}/f_{\omega}$, $f_{\phi NN}^{(1)}/f_{\phi}$,
$f_{\phi^, NN}^{(1)}/f_{\phi^,}$, $f_{\rho NN}^{(1)}/f_{\rho}$,
$f_{\rho^{,,,} NN}^{(1)}/f_{\rho^{,,,}}$,
$f_{\rho^{,,,,} NN}^{(1)}/f_{\rho^{,,,,}}$
$f_{\phi NN}^{(2)}/f_{\phi}$, $f_{\phi^, NN}^{(2)}/f_{\phi^,}$,
$f_{\rho^{,,,} NN}^{(2)}/f_{\rho^{,,,}}$,
$f_{\rho^{,,,,} NN}^{(2)}/f_{\rho^{,,,,}}$,
however, with the clear physical meaning. Not all of them are free.
For instance, owing to a simple reason that almost all (except for
$\rho^{,,,}$ and $\rho^{,,,,}$) considered resonances are situated in the
region $t_0<t<4m_N^2$, where no experimental informations on the nucleon
e.m. ff's exists up to now, one can not expect in a fitting procedure to be able
to determine their correct masses and widths. Therefore, for
$\omega$, $\phi$, $\omega^{,}$, $\omega^{,,}$, $\phi^,$,
$\rho$, $\rho^,$, and $\rho^{,,}$ they are fixed at the reliable
world averaged values
given by Review of Particle Physics \ct{29} and then investigated in the
framework of constructed model to be consistent with existing experimental
information on nucleon e.m. ff's. The parameters of $\rho^{,,,}$ are
taken from \ct{31}.

Thus we are left finally only with the following 16 free parameters
$t_{in}^{1s}$, $t_{in}^{1v}$, $t_{in}^{2s}$, $t_{in}^{2v}$,
$m_{\rho^{,,,,}}$, $\Gamma_{\rho^{,,,,}}$,
$f_{\omega NN}^{(1)}/f_{\omega}$, $f_{\phi NN}^{(1)}/f_{\phi}$,
$f_{\phi^, NN}^{(1)}/f_{\phi^,}$, $f_{\rho NN}^{(1)}/f_{\rho}$,
$f_{\rho^{,,,} NN}^{(1)}/f_{\rho^{,,,}}$,
$f_{\rho^{,,,,} NN}^{(1)}/f_{\rho^{,,,,}}$
$f_{\phi NN}^{(2)}/f_{\phi}$, $f_{\phi^, NN}^{(2)}/f_{\phi^,}$,
$f_{\rho^{,,,} NN}^{(2)}/f_{\rho^{,,,}}$,
$f_{\rho^{,,,,} NN}^{(2)}/f_{\rho^{,,,,}}$,
as the four effective inelastic thresholds are specific quantities
in the constructed
model, the fourth excited state of the $\rho$-meson is not identified
experimentally up to now
and at present there are no model independent and consistent values of the
$f_{\omega NN}^{(1)}/f_{\omega}$, $f_{\phi NN}^{(1)}/f_{\phi}$,
$f_{\phi^, NN}^{(1)}/f_{\phi^,}$, $f_{\rho NN}^{(1)}/f_{\rho}$,
$f_{\phi NN}^{(2)}/f_{\phi}$, $f_{\phi^, NN}^{(2)}/f_{\phi^,}$ coupling
constant ratios.

For their numerical evaluations we have collected 476 experimental points,
 mostly for the proton in the space-like region up to $t=-33\;GeV^2$,
however, including also new time like data of the proton \ct{13-18} and
the neutron \ct{19} above the $N\bar N$ threshold. The data have
been analyzed
by means of the relations (\ref{d5}) and (\ref{d30})-(\ref{d33}) by using the CERN program
MINUIT. The best description of them was achieved with $\chi^2/ndf=1.43$ and
the following values of free parameters
\begin{eqnarray}
\nonumber
&t_{in}^{1s}=2.6012\pm 0.6391\;GeV^2 &t_{in}^{1v}=3.5220\pm 0.0059\;GeV^2\\
\nonumber
&t_{in}^{2s}=2.7200\pm 0.6271\;GeV^2 &t_{in}^{2v}=3.6316\pm 0.6235\;GeV^2\\
\nonumber
&(f_{\omega NN}^{(1)}/f_{\omega})=1.1112\pm 0.0030 &(f_{\rho NN}^{(1)}/f_{\rho})=0.3843\pm 0.0043\\
\nonumber
&(f_{\phi NN}^{(1)}/f_{\phi})=-0.9389\pm 0.0056 &(f_{\rho^{,,,} NN}^{(1)}/f_{\rho^{,,,}})=-0.0840\pm 0.0008\\
&(f_{\phi^, NN}^{(1)}/f_{\phi^,})=-0.3255\pm 0.0047 &(f_{\rho^{,,,} NN}^{(2)}/f_{\rho^{,,,}})=0.0299\pm 0.0003
\label{d38}\\
\nonumber
&(f_{\phi NN}^{(2)}/f_{\phi})=-0.2659\pm 0.0287 &m_{\rho^{,,,,}}=2506\pm 38\;MeV\\
\nonumber
&(f_{\phi^, NN}^{(2)}/f_{\phi^,})=0.1190\pm 0.0032 &\Gamma_\ve=700\pm 179\;MeV\\
\nonumber
&\;&(f_{\rho^{,,,,} NN}^{(1)}/f_{\rho^{,,,,}})=0.0549\pm 0.0005\\
\nonumber
&\;&(f_{\rho^{,,,,} NN}^{(2)}/f_{\rho^{,,,,}})=-0.0103\pm 0.0001.
\end{eqnarray}

A compilation of the world nucleon ff data and their description by our
ten-resonance unitary and analytic model is graphically presented
in Figs. 1-4.
One can see from Fig. 4b that unlike the authors of the paper \ct{22}
we are able to describe FENICE time-like data on neutron \ct{19} quite well.
The same is valid also for the FERMILAB proton time-like data \ct{14,15} (see
Fig. 2b). The latter was possible to achieve by an introduction of a hypothetical
fourth excited state of the $\rho$(770)-meson, the parameters of which were
found in a fitting procedure of all existing data to be quite reasonable.
Its existence, however, has to be proved by an identification
also in other processes than only in $e^+e^-\to N\bar N$.

Of particular interest is a determination of the radii of the isoscalar and
isovector parts of the Dirac and Pauli ff's. They are given in Table 1, where
for a comparison results of the papers \ct{20} and
\ct{23} are presented too.
The corresponding proton and neutron radii are given in Table 2.
Here we would like to stress that we do not use in our model the neutron
charge radius to be determined very accurately by measuring the neutron-atom
scattering length \ct{32} as a constraint like in \ct{20} and it is a
prediction of the model to be $\langle r^2_{En}\rangle = -.097fm^2$.

In order to demonstrate explicitly substantial deviations from the dipole
fit in all channels and at the same time a violation of the
nucleon ff scaling, particularly at large momentum transfer, we show
in Figs. 5-8 ratios of appropriately  normalized electric and magnetic proton and neutron ff's in the space-like
region to the dipole formula $G_D(t)=(1-t/0.71)^{-2}$.

\section{Predictions of our unitary and analytic model of the nucleon e.m.
structure}

The unitary and analytic ten-resonance model of the nucleon e.m. structure
constructed in this paper represents a harmonious unification of all known
nucleon ff properties always into one analytic function, i.e. one smooth function
on the whole real axis, for every nucleon e.m. ff. As a result one
can believe
then the predicted behaviours of these nucleon e.m. ff's to be realistic
also outside the regions of existing experimental data.

Valuable is the predicted existence of the fourth excited state of
the $\rho$(770)-meson with resonance parameters $m_{\rho^{,,,,}}=2500\;MeV$
and $\Gamma_{\rho^{,,,,}}=700\;MeV$ without of which one could not achieve
a satisfactory description of the FENICE time-like neutron data \ct{19}
(see Fig. 4b) and
also eight FERMILAB proton points \ct{14,15} (Fig. 2b) at higher energies.

Taking into account the numerical results (\ref{d38}) of the parameters
and the transformed relations
(13)-(\ref{d16}) by means of the expressions (\ref{d34})-(\ref{d37}) into
the forms
\begin{eqnarray}
\nonumber {\rm I}.\;\;\; (f^{(1)}_{\omega^{,} NN}/f_{\omega^{,}})&=&\frac{1}{2}
 \frac{C^{1s}_{\omega^{,,}}}{C^{1s}_{\omega^{,,}}-C^{1s}_{\omega^{,}}}
 -(f^{(1)}_{\omega^{} NN}/f_{\omega^{}})
 \frac{C^{1s}_{\omega^{,,}}-C^{1s}_{\omega^{}}}
 {C^{1s}_{\omega^{,,}}-C^{1s}_{\omega^{,}}}-\\
&-&(f^{(1)}_{\phi^{} NN}/f_{\phi^{}})
 \frac{C^{1s}_{\omega^{,,}}-C^{1s}_{\phi^{}}}
 {C^{1s}_{\omega^{,,}}-C^{1s}_{\omega^{,}}}
 +(f^{(1)}_{\phi^{,} NN}/f_{\phi^{,}})
 \frac{C^{1s}_{\phi^{,}}-C^{1s}_{\omega^{,,}}}
 {C^{1s}_{\omega^{,,}}-C^{1s}_{\omega^{,}}}\label{d39}\\
\nonumber (f^{(1)}_{\omega^{,,} NN}/f_{\omega^{,,}})&=&-\frac{1}{2}
 \frac{C^{1s}_{\omega^{,}}}{C^{1s}_{\omega^{,,}}-C^{1s}_{\omega^{,}}}
 +(f^{(1)}_{\omega^{} NN}/f_{\omega^{}})
 \frac{C^{1s}_{\omega^{,}}-C^{1s}_{\omega^{}}}
 {C^{1s}_{\omega^{,,}}-C^{1s}_{\omega^{,}}}+\\
\nonumber &+&(f^{(1)}_{\phi^{} NN}/f_{\phi^{}})
 \frac{C^{1s}_{\omega^{,}}-C^{1s}_{\phi^{}}}
 {C^{1s}_{\omega^{,,}}-C^{1s}_{\omega^{,}}}
 -(f^{(1)}_{\phi^{,} NN}/f_{\phi^{,}})
 \frac{C^{1s}_{\phi^{,}}-C^{1s}_{\omega^{,}}}
 {C^{1s}_{\omega^{,,}}-C^{1s}_{\omega^{,}}}\\
\nonumber &\ &\ \\
\nonumber {\rm II}.\;\;\; (f^{(1)}_{\rho^{,} NN}/f_{\rho^{,}})&=&\frac{1}{2}
 \frac{C^{1v}_{\rho^{,,}}}{C^{1v}_{\rho^{,,}}-C^{1v}_{\rho^{,}}}
 -(f^{(1)}_{\rho^{} NN}/f_{\rho^{}})
 \frac{C^{1v}_{\rho^{,,}}-C^{1v}_{\rho^{}}}
 {C^{1v}_{\rho^{,,}}-C^{1v}_{\rho^{,}}}+\\
&+&(f^{(1)}_{\rho^{,,,} NN}/f_{\rho^{,,,}})
 \frac{C^{1v}_{\rho^{,,,}}-C^{1v}_{\rho^{,,}}}
 {C^{1v}_{\rho^{,,}}-C^{1v}_{\rho^{,}}}
 +(f^{(1)}_{\rho^{,,,,} NN}/f_{\rho^{,,,,}})
 \frac{C^{1v}_{\rho^{,,,,}}-C^{1v}_{\rho^{,,}}}
 {C^{1v}_{\rho^{,,}}-C^{1v}_{\rho^{,}}}\label{d40}\\
\nonumber (f^{(1)}_{\rho^{,,} NN}/f_{\rho^{,,}})&=&-\frac{1}{2}
 \frac{C^{1v}_{\rho^{,}}}{C^{1v}_{\rho^{,,}}-C^{1v}_{\rho^{,}}}
 +(f^{(1)}_{\rho^{} NN}/f_{\rho^{}})
 \frac{C^{1v}_{\rho^{,}}-C^{1v}_{\rho^{}}}
 {C^{1v}_{\rho^{,,}}-C^{1v}_{\rho^{,}}}-\\
\nonumber &-&(f^{(1)}_{\rho^{,,,} NN}/f_{\rho^{,,,}})
 \frac{C^{1v}_{\rho^{,,,}}-C^{1v}_{\rho^{,}}}
 {C^{1v}_{\rho^{,,}}-C^{1v}_{\rho^{,}}}
 -(f^{(1)}_{\rho^{,,,,} NN}/f_{\rho^{,,,,}})
 \frac{C^{1v}_{\rho^{,,,,}}-C^{1v}_{\rho^{,}}}
 {C^{1v}_{\rho^{,,}}-C^{1v}_{\rho^{,}}}\\
\nonumber &\ &\ \\
\nonumber {\rm III}.\;\;\; (f^{(2)}_{\omega^{} NN}/f_{\omega^{}})&=&\frac{1}{2}
 (\mu_p+\mu_n)\frac{C^{2s}_{\omega^{,,}}C^{2s}_{\omega^{,}}}
 {(C^{2s}_{\omega^{,,}}-C^{2s}_{\omega^{}})
  (C^{2s}_{\omega^{,}}-C^{2s}_{\omega^{}})}-\\
\nonumber &-&(f^{(2)}_{\phi^{} NN}/f_{\phi^{}})
 \frac{(C^{2s}_{\omega^{,,}}-C^{2s}_{\phi^{}})
       (C^{2s}_{\omega^{,}}-C^{2s}_{\phi^{}})}
       {(C^{2s}_{\omega^{,,}}-C^{2s}_{\omega^{}})
       (C^{2s}_{\omega^{,}}-C^{2s}_{\omega^{}})}-\\
\nonumber &-&(f^{(2)}_{\phi^{,} NN}/f_{\phi^{,}})
   \frac{(C^{2s}_{\phi^{,}}-C^{2s}_{\omega^{,,}})
       (C^{2s}_{\phi^{,}}-C^{2s}_{\omega^{,}})}
       {(C^{2s}_{\omega^{,,}}-C^{2s}_{\omega^{}})
       (C^{2s}_{\omega^{,}}-C^{2s}_{\omega^{}})}\\
\nonumber (f^{(2)}_{\omega^{,} NN}/f_{\omega^{,}})&=&-\frac{1}{2}
 (\mu_p+\mu_n)\frac{C^{2s}_{\omega^{,,}}C^{2s}_{\omega^{}}}
 {(C^{2s}_{\omega^{,,}}-C^{2s}_{\omega^{,}})
  (C^{2s}_{\omega^{,}}-C^{2s}_{\omega^{}})}-\\
&-&(f^{(2)}_{\phi^{} NN}/f_{\phi^{}})
 \frac{(C^{2s}_{\omega^{,,}}-C^{2s}_{\phi^{}})
       (C^{2s}_{\phi^{}}-C^{2s}_{\omega^{}})}
       {(C^{2s}_{\omega^{,,}}-C^{2s}_{\omega^{,}})
       (C^{2s}_{\omega^{,}}-C^{2s}_{\omega^{}})}+\label{d41}\\
\nonumber
       &+&(f^{(2)}_{\phi^{,} NN}/f_{\phi^{,}})
  \frac{(C^{2s}_{\phi^{,}}-C^{2s}_{\omega^{,,}})
       (C^{2s}_{\phi^{,}}-C^{2s}_{\omega^{}})}
       {(C^{2s}_{\omega^{,,}}-C^{2s}_{\omega^{,}})
       (C^{2s}_{\omega^{,}}-C^{2s}_{\omega^{}})}\\
\nonumber (f^{(2)}_{\omega^{,,} NN}/f_{\omega^{,,}})&=&\frac{1}{2}
 (\mu_p+\mu_n)\frac{C^{2s}_{\omega^{,}}C^{2s}_{\omega^{}}}
 {(C^{2s}_{\omega^{,,}}-C^{2s}_{\omega^{,}})
  (C^{2s}_{\omega^{,,}}-C^{2s}_{\omega^{}})}+\\
\nonumber
 &+&(f^{(2)}_{\phi^{} NN}/f_{\phi^{}})
       \frac{(C^{2s}_{\omega^{,}}-C^{2s}_{\phi^{}})
       (C^{2s}_{\phi^{}}-C^{2s}_{\omega^{}})}
       {(C^{2s}_{\omega^{,,}}-C^{2s}_{\omega^{,}})
       (C^{2s}_{\omega^{,,}}-C^{2s}_{\omega^{}})}-\\
\nonumber &-&  (f^{(2)}_{\phi^{,} NN}/f_{\phi^{,}})
  \frac{(C^{2s}_{\phi^{,}}-C^{2s}_{\omega^{,}})
       (C^{2s}_{\phi^{,}}-C^{2s}_{\omega^{}})}
       {(C^{2s}_{\omega^{,,}}-C^{2s}_{\omega^{,}})
       (C^{2s}_{\omega^{,,}}-C^{2s}_{\omega^{}})}\\
\nonumber &\ &\ \\
{\rm IV}.\;\;\; (f^{(2)}_{\rho^{} NN}/f_{\rho^{}})&=&\frac{1}{2}
 (\mu_p-\mu_n)\frac{C^{2v}_{\rho^{,,}}C^{2v}_{\rho^{,}}}
 {(C^{2v}_{\rho^{,,}}-C^{2v}_{\rho^{}})
  (C^{2v}_{\rho^{,}}-C^{2v}_{\rho^{}})}-\label{d42}\\
\nonumber &-&(f^{(2)}_{\rho^{,,,} NN}/f_{\rho^{,,,}})
 \frac{(C^{2v}_{\rho^{,,,}}-C^{2v}_{\rho^{,,}})
       (C^{2v}_{\rho^{,,,}}-C^{2v}_{\rho^{,}})}
       {(C^{2v}_{\rho^{,,}}-C^{2v}_{\rho^{}})
       (C^{2v}_{\rho^{,}}-C^{2v}_{\rho^{}})}-\\
\nonumber &-&
  (f^{(2)}_{\rho^{,,,,} NN}/f_{\rho^{,,,,}})
  \frac{(C^{2v}_{\rho^{,,,,}}-C^{2v}_{\rho^{,,}})
       (C^{2v}_{\rho^{,,,,}}-C^{2v}_{\rho^{,}})}
       {(C^{2v}_{\rho^{,,}}-C^{2v}_{\rho^{}})
       (C^{2v}_{\rho^{,}}-C^{2v}_{\rho^{}})}\\
\nonumber (f^{(2)}_{\rho^{,} NN}/f_{\rho^{,}})&=&-\frac{1}{2}
 (\mu_p-\mu_n)\frac{C^{2v}_{\rho^{,,}}C^{2v}_{\rho^{}}}
 {(C^{2v}_{\rho^{,,}}-C^{2v}_{\rho^{,}})
  (C^{2v}_{\rho^{,}}-C^{2v}_{\rho^{}})}+\\
\nonumber &+&(f^{(2)}_{\rho^{,,,} NN}/f_{\rho^{,,,}})
 \frac{(C^{2v}_{\rho^{,,,}}-C^{2v}_{\rho^{,,}})
       (C^{2v}_{\rho^{,,,}}-C^{2v}_{\rho^{}})}
       {(C^{2v}_{\rho^{,,}}-C^{2v}_{\rho^{,}})
       (C^{2v}_{\rho^{,}}-C^{2v}_{\rho^{}})}+\\
\nonumber &+&
  (f^{(2)}_{\rho^{,,,,} NN}/f_{\rho^{,,,,}})
  \frac{(C^{2v}_{\rho^{,,,}}-C^{2v}_{\rho^{,,}})
       (C^{2v}_{\rho^{,,,,}}-C^{2v}_{\rho^{}})}
       {(C^{2v}_{\rho^{,,}}-C^{2v}_{\rho^{,}})
       (C^{2v}_{\rho^{,}}-C^{2v}_{\rho^{}})}\\
\nonumber (f^{(2)}_{\rho^{,,} NN}/f_{\rho^{,,}})&=&\frac{1}{2}
 (\mu_p-\mu_n)\frac{C^{2v}_{\rho^{,}}C^{2v}_{\rho^{}}}
 {(C^{2v}_{\rho^{,,}}-C^{2v}_{\rho^{,}})
  (C^{2v}_{\rho^{,,}}-C^{2v}_{\rho^{}})}-\\
\nonumber &-&(f^{(2)}_{\rho^{,,,} NN}/f_{\rho^{,,,}})
 \frac{(C^{2v}_{\rho^{,,,}}-C^{2v}_{\rho^{,}})
       (C^{2v}_{\rho^{,,,}}-C^{2v}_{\rho^{}})}
       {(C^{2v}_{\rho^{,,}}-C^{2v}_{\rho^{,}})
       (C^{2v}_{\rho^{,,}}-C^{2v}_{\rho^{}})}-\\
\nonumber &-&
  (f^{(2)}_{\rho^{,,,,} NN}/f_{\rho^{,,,,}})
  \frac{(C^{2v}_{\rho^{,,,,}}-C^{2v}_{\rho^{,}})
       (C^{2v}_{\rho^{,,,,}}-C^{2v}_{\rho^{}})}
       {(C^{2v}_{\rho^{,,}}-C^{2v}_{\rho^{,}})
       (C^{2v}_{\rho^{,,}}-C^{2v}_{\rho^{}})}
\end{eqnarray}
the following additional coupling constant ratio values are predicted
\begin{eqnarray}
\nonumber &(f^{(1)}_{\omega^{,}NN}/f_{\omega^{,}})=0.5045
&(f^{(1)}_{\rho^{,}NN}/f_{\rho^{,}})=0.7647\\
\nonumber &(f^{(1)}_{\omega^{,,}NN}/f_{\omega^{,,}})=0.1482
&(f^{(1)}_{\rho^{,,}NN}/f_{\rho^{,,}})=-0.6199\\
&(f^{(2)}_{\omega^{}NN}/f_{\omega^{}})=0.1712
&(f^{(2)}_{\rho^{}NN}/f_{\rho^{}})=3.0530\label{d43}\\
\nonumber &(f^{(2)}_{\omega^{,}NN}/f_{\omega^{,}})=-.02455
&(f^{(2)}_{\rho^{,}NN}/f_{\rho^{,}})=-1.6790\\
\nonumber &(f^{(2)}_{\omega^{,,}NN}/f_{\omega^{,,}})=-.05992
&(f^{(2)}_{\rho^{,,}NN}/f_{\rho^{,,}})=1.0040.
\end{eqnarray}
By a combination of the numerical results in (\ref{d38}) and (\ref{d43})
one predicts
the tensor-to-vector coupling constant ratios. In particular for $\rho$(770)
and $\omega$(1020) one obtains
\begin{eqnarray}
\nonumber &K_{\rho}=(f^{(2)}_{\rho NN}/f_{\rho})/(f^{(1)}_{\rho NN}/f_{\rho})=7.945\\
&K_{\omega}=(f^{(2)}_{\omega NN}/f_{\omega})/(f^{(1)}_{\omega NN}/f_{\omega})=0.154.
\label{d44}\end{eqnarray}
Taking into account the strict VMD model giving $K_\rho=\mu_p-\mu_n=3.71$,
$K_\omega=\mu_p+\mu_n=-0.12$ and evaluations of other authors \ct{20,23}
$K_\rho$ to be large ($\sim 6$) and $K_\omega$ to be small ($\sim 0$), our
results seem to be quite reasonable. Other tensor-to-vector coupling ratios
are:
\begin{eqnarray}
\nonumber &K_\phi=0.283,\; K_{\omega^,}=-0.049,\; K_{\omega^{,,}}=-0.404,\;
K_{\phi^,}=-0.366,\\
&K_{\rho^,}=-2.196\;,K_{\rho^{,,}}=-1.620,\;K_{\rho^{,,,}}=-0.356,\;K_{\rho^{,,,,}}=-0.188.\label{d45}
\end{eqnarray}

The universal vector meson coupling constants $f_s$ and $f_v$ are determined
from the widths of the leptonic decays, i.e.
\begin{equation}
\frac{f_v^2}{4\pi}=\frac{\alpha^2}{3}\frac{m_v}{\Gamma(\nu\to e^+e^-)}.
\label{d46}\end{equation}
Then numerical values
\begin{equation}
f_\rho=5.0320\pm 0.1089; \; f_\omega= 17.0499\pm 0.2990;\; f_\phi=-12.8832\pm 0.0824
\label{d47}\end{equation}
are obtained from the corresponding world averaged lepton widths \ct{29}
and the universal $\omega^,-$, $\omega^{,,}-$ and $\rho^,-$, $\rho^{,,}-$
meson coupling constants
\begin{eqnarray}
f_{\omega^,}=47.6022\pm 7.5026;\;f_{\omega^{,,}}=48.3778\pm 7.5026
\label{d48}\end{eqnarray}
and
\begin{eqnarray}
f_{\rho^,}=13.6491\pm 0.9521;\;f_{\rho^{,,}}=22.4020\pm 2.2728
\label{d49}\end{eqnarray}
have been found from the lepton widths estimated by Donnachie and Clegg \ct{33}.

As a result, the following numerical values of the corresponding coupling
constants are predicted
\begin{eqnarray}
\nonumber &f^{(1)}_{\omega^{}NN}=18.9527; &f^{(1)}_{\rho^{}NN}=1.9335; \\
\nonumber &f^{(1)}_{\phi^{}NN}= 12.0956; &f^{(1)}_{\rho^{,}NN}=10.4375; \\
\nonumber &f^{(1)}_{\omega^{,}NN}=24.0153; &f^{(1)}_{\rho^{,,}NN}=-13.8870; \\
\nonumber &f^{(1)}_{\omega^{,,}NN}=7.1696; & \\
          &\label{d50}\\
\nonumber &f^{(2)}_{\omega^{}NN}=2.9189; &f^{(2)}_{\rho^{}NN}=15.3627; \\
\nonumber &f^{(2)}_{\phi^{}NN}= 3.4251; &f^{(2)}_{\rho^{,}NN}=-22.9168; \\
\nonumber &f^{(2)}_{\omega^{,}NN}=-1.1686; &f^{(2)}_{\rho^{,,}NN}= 22.4916;\\
\nonumber &f^{(2)}_{\omega^{,,}NN}=-2.8988. &
\end{eqnarray}

Their squares divided by $4\pi$ are presented in Table 3 and Table 4,
where for a
comparison also values obtained by other authors are shown.

One can immediately notice  large value of $f^{(1,2)}_{\phi NN}$ coupling
constants which may indicate a violation of OZI rule \ct{34}.

By using the numerical values (\ref{d50}) one can predict the
$\omega-\phi$ mixing
angle, employing the relation
\begin{equation}
\frac{\sqrt{3}}{\cos\vartheta}\frac{f^{(1)}_{\rho NN}}{f^{(1)}_{\omega NN}}
-\tan\vartheta=\frac{f^{(1)}_{\phi NN}}{f^{(1)}_{\omega NN}}.\label{d51}
\end{equation}
It takes the value $\vartheta=0.7175$, which is very near to the ideal mixing.

Our model predicts the neutron charge radius $r_E^n$ to be negative automatically
and we are not in need to secure this phenomenon by a constraint following
from the low-energy neutron-atom scattering results like in \ct{20}.

Nevertheless, the most important predictions of our unitary and analytic
model of the nucleon e.m. structure are the isovector spectral function
behaviours (see Fig.9) to be consistent with predictions of H\H{o}hler and
Pietarinen \ct{27} and Mergel, Mei\ss ner and Drechsel \ct{20}, which have been
carried out on the base of the Frazer and Fulco \ct{35}
unitarity relation by using
the pion e.m. ff $F_\pi(t)$ and the $P$-wave $\pi\pi\to N\bar N$
partial wave amplitudes obtained by an analytic continuation
of the experimental
information on $\pi N$-scattering into the unphysical region.

The method of our prediction of the latter consist in the following. The ten-resonance
unitary and analytic model of nucleon e.m. structure constructed in
this paper contains an explicit two-pion continuum contribution given
by the unitary cut starting from $t=4m_\pi^2$, from where just isovector
spectral function start to be different from zero. Then despite of the
fact that the unstable $\rho$-meson is taken into account as
complex conjugate pairs of  poles shifted
from the real axis into the complex plane on the second and third
Riemann sheet of
the four-sheeted Riemann surface, the model predicts the strong enhancement
on the left wing of the $\rho$(770) resonance in the isovector spectral
functions
automatically. And just agreement of our predictions with those obtained by
means of the Frazer and Fulco unitarity relation convinces us that our model
constructed in this paper is really unitary.

Another success of the presented model is a prediction of isoscalar nucleon
spectral
function behaviours (see Fig.10) for the first time as the model contains
an explicit three-pion continuum contribution given by the unitary cut
starting from $t=9m_\pi^2$, from where just isoscalar spectral functions
start to be different from zero.

\section{Conclusions}

We have constructed unitary and analytic ten-resonance (5 isoscalars and 5
isovectors) model of the nucleon e.m. structure which represents a harmonious
unification of all known nucleon ff properties, like analyticity, reality
condition, experimental fact of creation of vector-meson resonances in
electron-positron annihilation processes, normalization and
the asymptotic behaviour as
predicted for nucleon e.m. ff's by the quark model of hadrons. It depends
only on parameters with clear physical meaning. They are four effective
square-root branch points, representing contribution of all other higher
thresholds given by the unitarity condition, the mass and width of the
hypothetical fourth excited state of $\rho$(770)-meson and coupling constants
of some resonances under consideration. They all are numerically evaluated
by an analysis of all existing space-like and time-like nucleon ff data.

We would like to note that by means of our model presented in this paper
we have described all existing nucleon ff data for the first time, including
also FENICE neutron time-like data and FERMILAB proton eight points
at higher energies. In the latter  an existence of the
$\rho^{,,,,}$(2500) resonance with parameters $m_{\rho^{,,,,}}=2500\;MeV$ and
$\Gamma_{\rho^{,,,,}}=700\;MeV$ is crucial. So, there is challenge to experimental
physicists to confirm an existence of this resonance also in other processes than
$e^+e^-\to N\bar N$.

Our unitary and analytic ten-resonance nucleon ff model gives a lot of
reasonable predictions. However, the most important among them are isoscalar
and isovector spectral function behaviours, which coincide also with
predictions obtained in the framework of heavy baryon chiral perturbation
theory \ct{24}.

 The work was in part supported   by Slovak Grant Agency for Sciences,
Grant No. 2/5085/99(S.D.) and Grant No. 1/4301/99(A.Z.D).

One of us (S.D.) would like to thank Prof. S. Randjbar-Daemi for the
hospitality at the ICTP Trieste where this work was completed.

\newpage
{\bf Figure captions}

\begin{itemize}
\item[Figs. 1a-2b]
A simultanious optimal fit of all existing data on
proton e.m.
ff's in the space-like and time-like regions by the unitary and analytic ten resonance
model of the proton e.m. structure, represented by expressions (5) and
(30)-(33).
\item[Figs. 3a-4b]
 A simultanious optimal fit of all existing data on neutron e.m.
ff's in the space-like and time-like regions by the unitary and analytic ten resonance
model of the neutron e.m. structure, represented by expressions (5) and
(30)-(33).
\item[Figs. 5-8]
 Ratios of appropriately normalized electric and magnetic
proton and neutron ff's in the space-like region
to the dipole formula.
\item[Fig. 9]
Predicted behaviours of the isovector spectral functions by 
the ten-resonance unitary and analytic model of the nucleon e.m. structure.
\item[Fig. 10]: 
Predicted behaviours of the corresponding isoscalar spectral functions.
\end{itemize}

\newpage
\begin{table}[t]
\caption{\bf Dirac and Pauli isoscalar and isovector radii
 of nucleons}
\begin{center}
\begin{tabular}{|c|c|c|c|c|} \hline
&$r^{(s)}_1[fm]$ & $r^{(s)}_2[fm]$ & $r^{(v)}_1[fm]$ & $r^{(v)}_2[fm]$\\ \hline \hline
our results&0.771&0.283&0.733&0.898 \\ \hline
Ref. [20]&0.782&0.845&0.765&0.893 \\ \hline
Ref. [23]&0.767&0.837&0.759&0.863 \\ \hline
\end{tabular}
\end{center}
\end{table}

\begin{table}[t]
\caption{\bf Electric and magnetic, and also Dirac and Pauli radii
of the proton and neutron}
\begin{center}
\begin{tabular}{|c|c|c|c|c|c|c|} \hline
&$r^{p}_E[fm]$ & $r^{p}_M[fm]$ & $r^{n}_M[fm]$ & $r^{p}_1[fm]$ & $r^{p}_2[fm]$ & $r^{n}_2[fm]$\\ \hline \hline
our results&0.827&0.860&0.891&0.752&0.914&0.883 \\ \hline
Ref. [20]&0.847&0.836&0.889&0.774&0.894&0.893 \\ \hline
Ref. [23]&0.836&0.843&0.840&0.761&0.883&0.876\\ \hline
\end{tabular}
\end{center}
\end{table}

\begin{table}[t]
\caption{\bf Coupling constants of the isoscalar vector mesons to nucleons}
\begin{center}
\begin{tabular}{|c|c|c|c|c|} \hline
 &$f_{\omega NN}^{(1)2}/4\pi$&$f_{\phi NN}^{(1)2}/4\pi$ & $f_{\omega' NN}^{(1)2}/4\pi$ & $f_{\omega'' NN}^{(1)2}/4\pi$ \\
 \hline \hline
our results&28.58&11.64&45.89&4.09  \\ \hline
Ref.[20]&34.6&6.7&--&--       \\ \hline
Ref. [23]&24.0&5.1&--&--    \\ \hline
\end{tabular}
\begin{tabular}{|c|c|c|c|c|c|} \hline
 & $f_{\omega NN}^{(2)2}/4\pi$ & $f_{\phi NN}^{(2)2}/4\pi$ & $f_{\omega' NN}^{(2)2}/4\pi$ & $f_{\omega'' NN}^{(2)2}/4\pi$\\
 \hline \hline
our results&0.67&0.93&0.11&0.67  \\ \hline
Ref.[20]&0.9&0.3&--&--        \\ \hline
Ref. [23]&--&0.2&--&--    \\ \hline
\end{tabular}
\end{center}
\end{table}

\begin{table}[t]
\caption{\bf Coupling constants of the isovector vector mesons to nucleons}
\begin{center}
\begin{tabular}{|c|c|c|c|} \hline
 &$f_{\rho NN}^{(1)2}/4\pi$&$f_{\rho' NN}^{(1)2}/4\pi$ & $f_{\rho'' NN}^{(1)2}/4\pi$  \\
 \hline \hline
our results&0.30&8.67&15.35 \\ \hline
Ref.[20]&--&40.27&793.53       \\ \hline
Ref. [23]&0.55&--&--    \\ \hline
\end{tabular}
\begin{tabular}{|c|c|c|c|c|c|} \hline
 & $f_{\rho NN}^{(2)2}/4\pi$ & $f_{\rho' NN}^{(2)2}/4\pi$ & $f_{\rho'' NN}^{(2)2}/4\pi$  \\
 \hline \hline
our results&18.78&41.79&40.26  \\ \hline
Ref.[20]&--&143.97&304.07        \\ \hline
Ref. [23]&24.0&11.5&--    \\ \hline
\end{tabular}
\end{center}
\end{table}

\newpage
\begin{figure}[t]
\psfig{figure=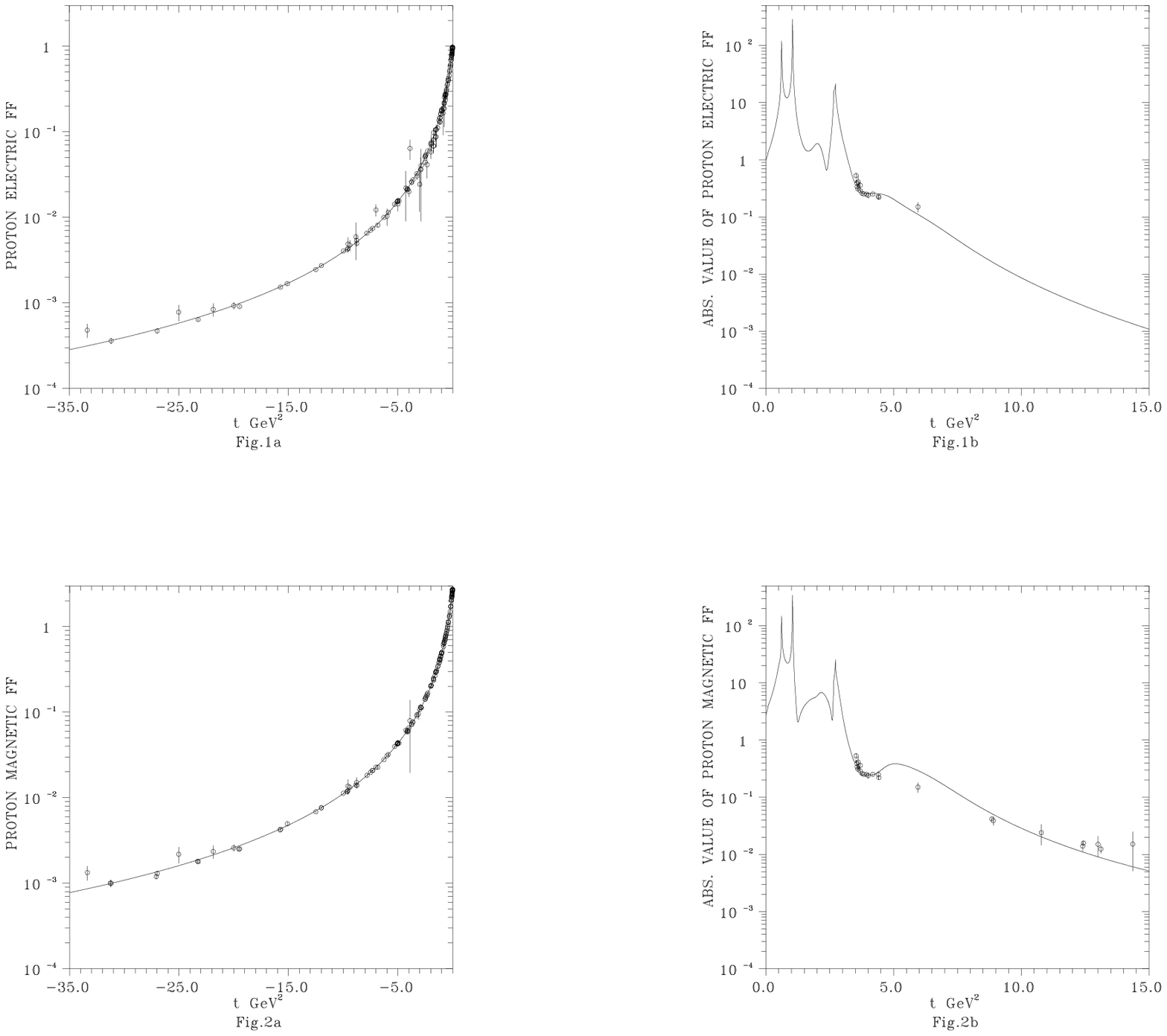,width=16cm}
\end{figure}
\newpage
\begin{figure}[t]
\psfig{figure=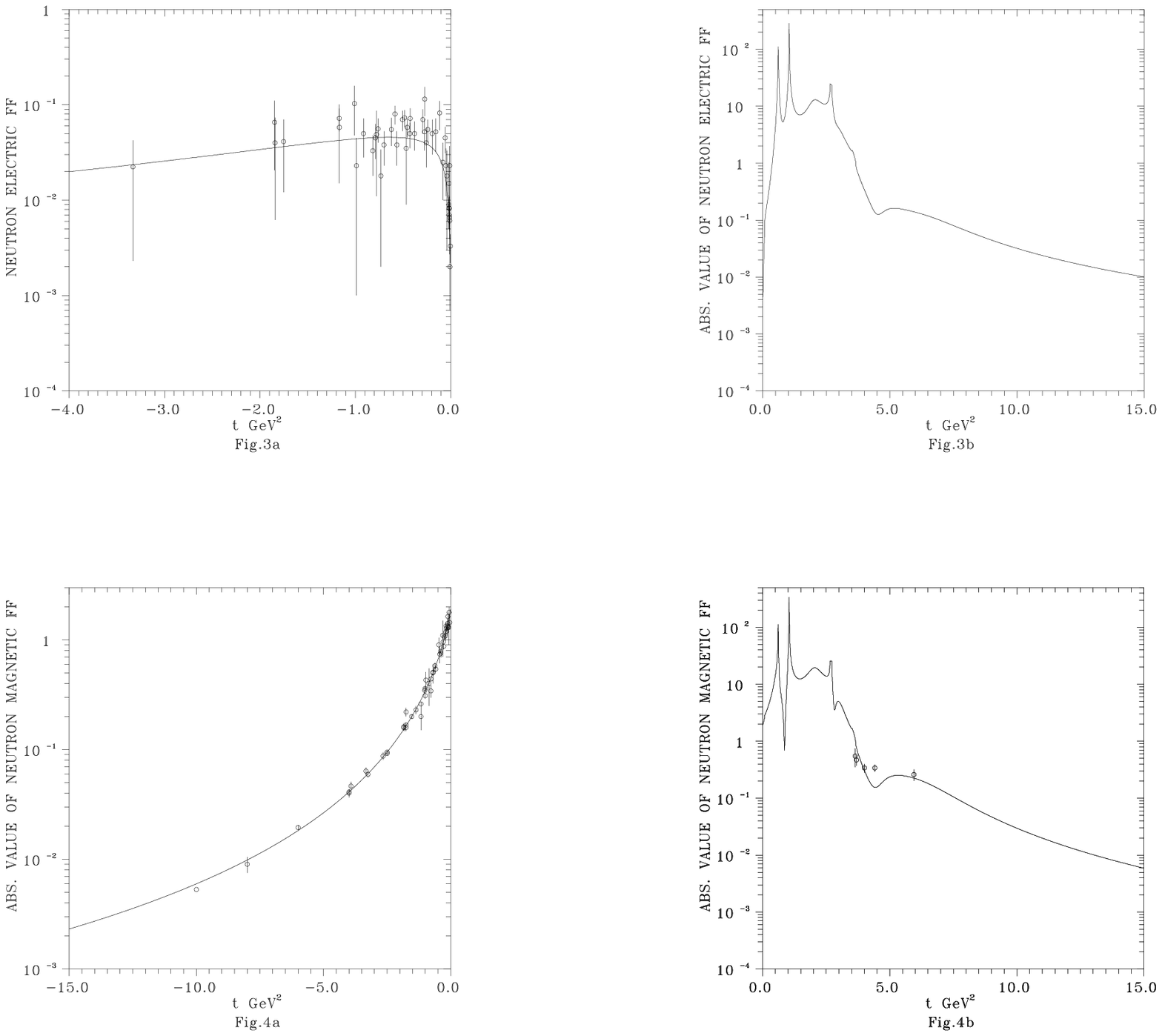,width=16cm}
\end{figure}
\newpage
\begin{figure}[thb]
\psfig{figure=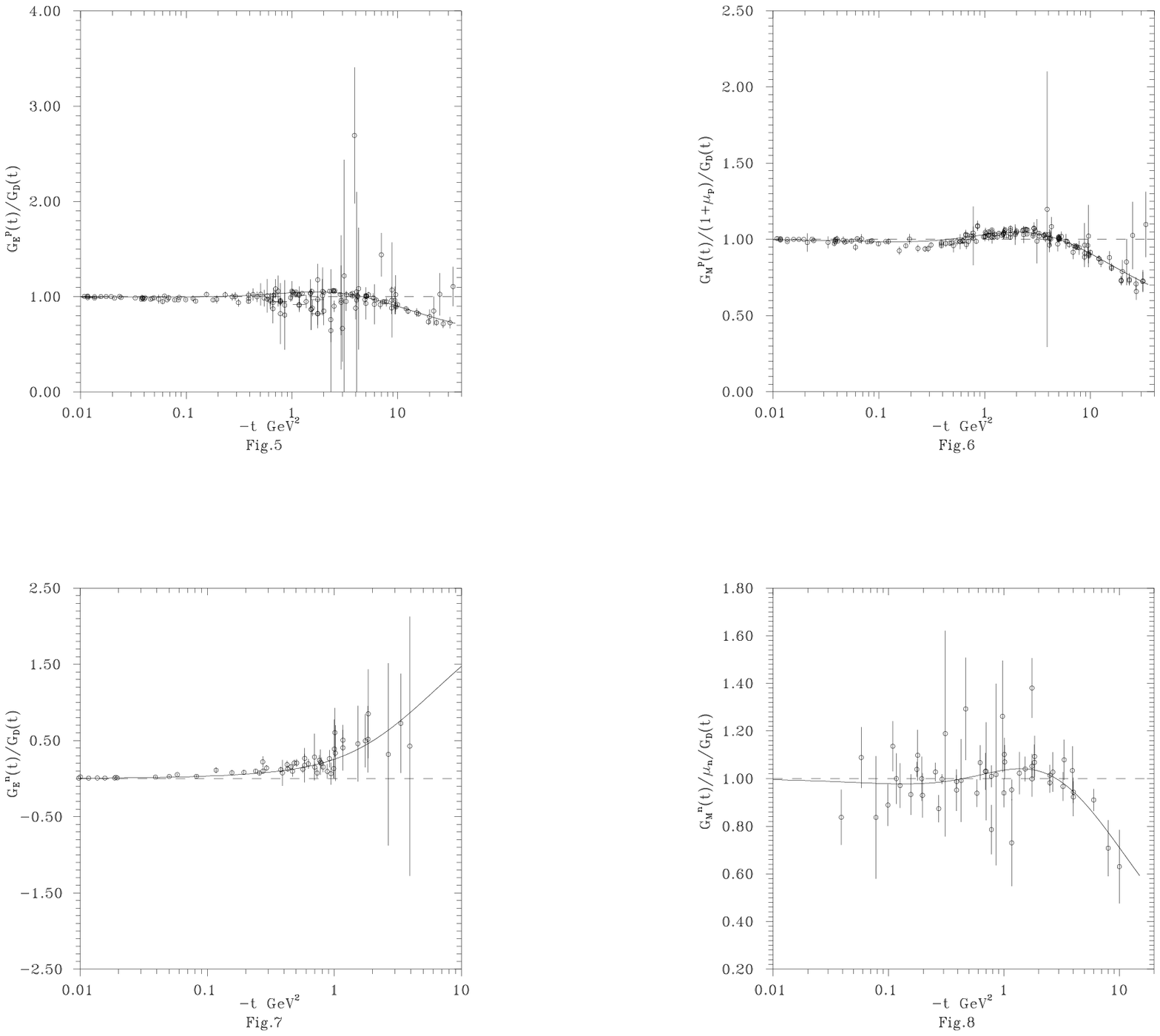,width=16cm}
\end{figure}

\newpage
\begin{figure}[t]
\hspace{2cm}\psfig{figure=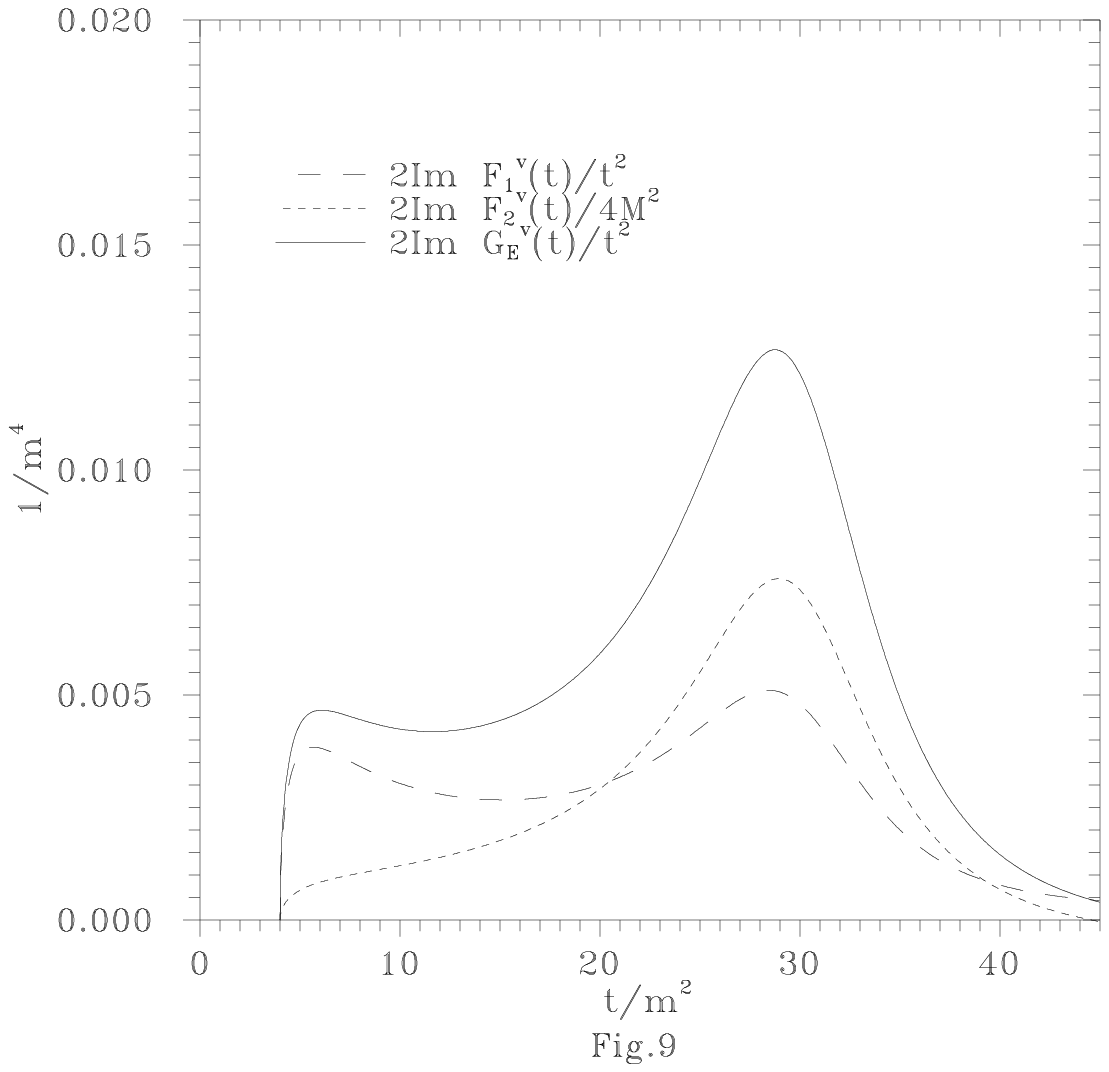,width=9.6cm}
\end{figure}

\begin{figure}[b]
\hspace*{2cm}\psfig{figure=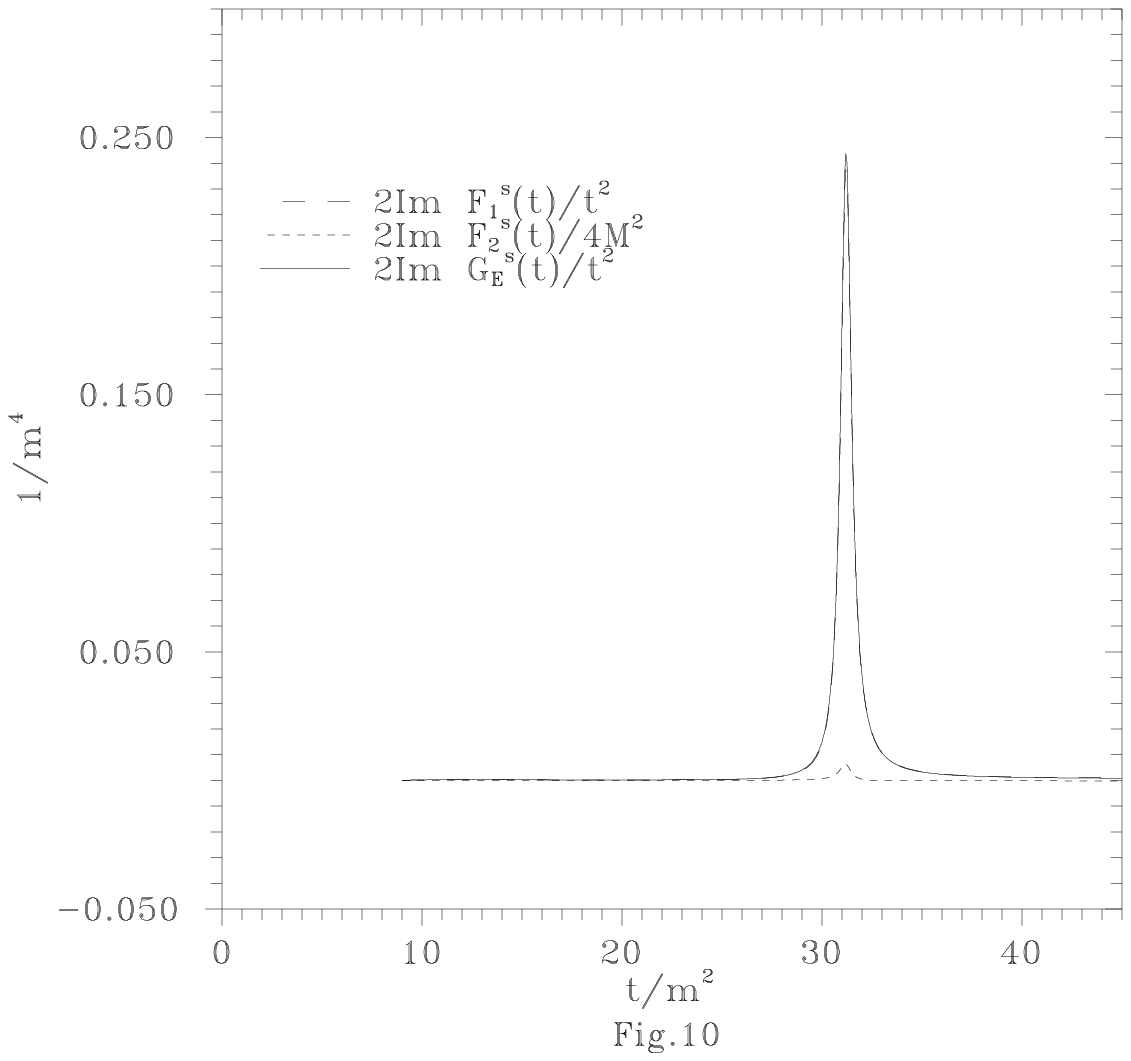,width=9.6cm}
\end{figure}


\begin{thebibliography}{99}
\bibitem{1}
S. Dubni\v cka, Nuovo Cim. {\bf A100}, (1988) 1.
\bibitem{2}
R. C. Walker et al, Phys. Rev {\bf D49}, (1994) 5671.
\bibitem{3}
L. Andivahis et al, Phys. Rev. {\bf D50}, (1994) 5491.
\bibitem{4}
A. F. Sill et al, Phys. Rev. {\bf D48}, (1993) 29.
\bibitem{5}
A. Lung et al, Phys. Rev. Lett. {\bf 70}, (1993) 718.
\bibitem{6}
S. Rock et al, Phys. Rev. {\bf D46}, (1992) 24.
\bibitem{7}
P. Markowitz et al, Phys. Rev. {\bf C48}, (1993) 5.
\bibitem{8}
T. Eden et al, Phys. Rev. {\bf C50}, (1994) 1749.
\bibitem{9}
M. Mayerhoff et al, Phys. Lett. {\bf 327B}, (1994) 201.
\bibitem{10}
S. Platchkov et al, Nucl. Phys. {\bf 510A}, (1990) 740.
\bibitem{11}
G. Bassompierre et al, Nuovo Cimento {\bf A73}, (1983) 347.
\bibitem{12}
B. Delcourt et al, Phys. Lett. {\bf 86B}, (1979) 395.
\bibitem{13}
D. Bisello et al, Nucl. Phys. {\bf B224}, (1893) 379.
\bibitem{14}
T.A. Armstrong et al, Phys. Rev. Lett. {\bf 70}, (1993) 1212.
\bibitem{15}
M. Ambrogioni et al, Phys. Rev. {\bf D60} (1999) 032002-1.
\bibitem{16}
D. Bisello et al, J. Phys. {\bf C48}, (1990) 23.
\bibitem{17}
G. Bardin et al, Nucl. Phys. {\bf B411}, (1994) 3.
\bibitem{18}
A. Antonelli et al, Phys. Lett. {\bf 334B}, (1994) 431.
\bibitem{19}
A. Antonelli et al, Nucl. Phys. {\bf B517}, (1998) 3. 
\bibitem{20}
P. Mergell, U.-G. Meissner and D. Drechsel, Nucl. Phys. {\bf A596}, (1996) 367.
\bibitem{21}
S. Furiuchi and D. Watanabe, Nuovo Cimento {\bf A110}, (1997) 577.
\bibitem{22}
H.-W. Hammer, Ulf-G. Meissner and D. Drechsel, Phys. Lett. {\bf 385B}, (1996) 343.
\bibitem{23}
G. H\H{o}hler et al, Nucl. Phys. {\bf B114}, (1976) 505.
\bibitem{24}
V. Bernard, N. Kaiser and Ulf-G. Meissner, Nucl. Phys. {\bf A611}, (1996) 429.
\bibitem{25}
S. Dubni\v cka, Nuovo Cimento {\bf A103}, (1990) 1417.
\bibitem{26}
S. Dubni\v cka, A. Z. Dubni\v ckova and P. Stri\v zenec, Nuovo Cimento {\bf A106}, (1993) 1253.
\bibitem{27}
G. H\H{o}hler and E. Pietarinen, Phys. Lett. {\bf 53B}, (1975) 471.
\bibitem{28}
S. J. Brodsky and P. G. Lepage, Phys. Rev. {\bf D22}, (1980) 2157.
\bibitem{29}
C. Caso et al, European Physical Journal {\bf 3} (1998) 1.
\bibitem{30}
S. I. Bilenkaya, S. Dubni\v cka, A. Z. Dubni\v ckova and P. Stri\v zenec,
Nuovo Cimento {\bf A105}, (1992) 1421.
\bibitem{31}
M. E. Biagini, S. Dubni\v cka, E. Etim and P. Kola\v r, Nuovo Cimento {\bf A104}, (1991) 363.
\bibitem{32}
S. Kopecky et al, Phys. Rev. Lett. {\bf 74} (1995) 2427.
\bibitem{33}
A. Donnachie and A.B. Clegg, Z. Phys. {\bf C42} (1989) 663.
\bibitem{34}
H. Genz and G. H\H{o}hler, Phys. Lett. {\bf 61B} (1976) 389.
\bibitem{35}
W.R. Frazer and J.R. Fulco, Phys. Rev. {\bf 117} (1960) 1603,1609.
\end{thebibliography}
\end{document}